\documentclass[journal,comsoc]{IEEEtran}

\author{Pinyarash~Pinyoanuntapong*,~\IEEEmembership{Member,~IEEE,}
Prabhu~Janakaraj*,~\IEEEmembership{Member,~IEEE,}
Ravikumar~Balakrishnan,~\IEEEmembership{Member,~IEEE,}
Minwoo~Lee,~\IEEEmembership{Member,~IEEE,}
Chen~Chen,~\IEEEmembership{Member,~IEEE,}
and~Pu~Wang,~\IEEEmembership{Member,~IEEE,}

\thanks{P. Pinyoanuntapong, P. Janakaraj, M. Lee and P. Wang are with the Department of Computer Science, University of North Carolina at Charlotte, Charlotte, NC, 28223 USA e-mail: {ppinyoan, pjanakar, Minwoo.Lee, pu.wang}@uncc.edu}
\thanks{R. Balakrishanan is with Intel Labs, USA}
\thanks{C. Chen is with the Center for Research in Computer Vision, University of Central Florida, Orlando, FL, 32816 USA e-mail: chen.chen@ucf.edu}
\thanks{The preliminary version of this paper was presented at 22nd IEEE International Workshop on Signal Processing Advances in Wireless Communications (SPAWC 2020) \cite{fedair} }
\thanks{* Authors contributed equally to this work}
 \thanks{This work is funded by Intel/NSF joint grant 2003198 and NSF 2008447}
 \thanks{Submitted to Journal of Computer Network}
 }

\usepackage[T1]{fontenc}

\usepackage{url}

\usepackage{multicol}
\usepackage{graphicx} 
\usepackage{mathptmx} 
\usepackage{times} 
\usepackage{amsmath} 
\usepackage{bm}
\usepackage{subfigure}
\usepackage{comment}
\usepackage{booktabs}
\usepackage{colortbl}
\usepackage[center]{caption}
\usepackage{float}
\usepackage{algorithm,algorithmic}

\usepackage{amsmath, amssymb}
\usepackage{graphicx}
\graphicspath{ {fig/} }
\usepackage{textcomp}
\usepackage{xcolor}
\usepackage{boxhandler}

\usepackage{cite}

\begin{document}

\title{EdgeML: Towards Network-Accelerated Federated Learning over Wireless Edge}
\maketitle

\begin{abstract}
 Federated learning (FL) is a distributed machine learning technology for next-generation AI systems that allows a number of workers, i.e., edge devices, collaboratively learn a shared global model while keeping their data locally to prevent privacy leakage. Enabling FL over wireless multi-hop networks can democratize AI and make it accessible in a cost-effective manner. However, the  noisy bandwidth-limited multi-hop wireless connections can lead to delayed and nomadic model updates, which significantly slows down the FL convergence speed. To address such challenges, this paper aims to accelerate FL convergence over wireless edge by optimizing the multi-hop federated networking performance. In particular, the FL convergence optimization problem is formulated as a  Markov decision process (MDP).  To solve such MDP,  multi-agent reinforcement learning (MA-RL) algorithms along with domain-specific action space refining schemes are developed, which online learn the delay-minimum forwarding paths to minimize the model exchange latency between the edge devices (i.e., workers) and the remote server. To validate the proposed solutions, FedEdge is developed and implemented, which is the first experimental framework in the literature for FL over multi-hop wireless edge computing networks. FedEdge allows us to fast prototype, deploy, and evaluate novel FL algorithms along with RL-based system optimization methods in real wireless devices. Moreover, a physical experimental testbed is implemented by customizing the widely adopted Linux wireless routers and ML computing nodes. Such testbed can provide valuable insights into the practical performance of FL in the field. Finally, our experimentation results on the testbed show that the proposed network-accelerated FL system can practically and significantly improve FL convergence speed, compared to the FL system empowered by the production-grade commercially-available wireless networking protocol, BATMAN-Adv.

%
%
%
\end{abstract}

\section{Introduction:}
Distributed machine learning, specifically federated learning (FL), has been envisioned as a key technology for enabling next-generation AI at scale. FL significantly reduces privacy risks and communication costs, which are critical in modern AI systems. FL allows workers (i.e., edge devices) to collaboratively learn a global model and maintains the locality of data to reduce the privacy vulnerability. The workers only need to send their local model updates to the server, which aggregates these updates to continuously improve the shared global model. FL can greatly reduce the required number of communication rounds for model convergence by increasing computation parallelization, where more edge devices are involved as the workers, and by increasing local computation, where the worker performs multiple iterations of model updates before sending the updated model to the server. Through FL, edge devices can still learn much more accurate models with small local datasets. As a result, FL has demonstrated its success for a variety of applications, such as on-device item ranking, content suggestions for on-device keyboards, and next word prediction \cite{FL}.

 \begin{figure}
	\centering
	\includegraphics[width=2.5in]{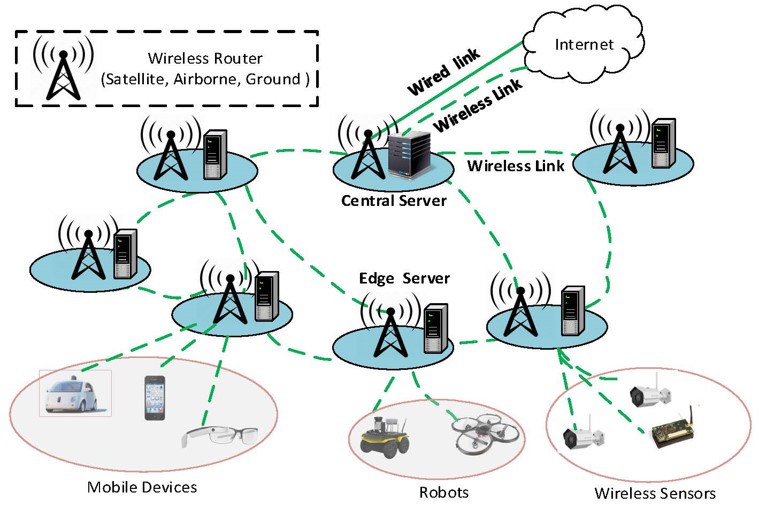}
	\caption{Multi-Hop Wireless Edge Computing Networks}
	\label{fig:MHFL}
\end{figure}

Recently, FL systems over edge computing networks have received increasing attentions. With single-hop wireless connections, edge devices can quickly reach the FL servers co-located with cellular base stations \cite{wirelessFL1,wirelessFL2,wirelessFL3,wirelessFL4,wirelessFL5}. Different from cellular systems with high deployment and operational costs, wireless multi-hop networks, consisting of a mesh of interconnected wireless routers, have been widely exploited to build cost-efficient communication backbones, including \textit{wireless community mesh networks} \cite{community2011,community} (e.g., Detroit digital stewards network \cite{Detroit}, Brooklyn Redhook WiFi \cite{redhookwifi}, NYC mesh \cite{nycmesh}, and Germany Freifunk network \cite{Germany}), \textit{high-speed urban networks} (e.g., Facebook Terragraph network \cite{terragraph} and London small cell mesh network \cite{london}), \textit{global wireless Internet infrastructures} (e.g., SpaceX Starlink satellite constellation \cite{starlink} and Google Loon balloon network \cite{loon}), \textit{battlefield networks} (e.g., rajant kinetic battlefield mesh networks \cite{battle}), and \textit{public safety/disaster recuse networks} \cite{safty}. The edge computing devices interconnected by the wireless multi-hop network constitute the multi-hop wireless edge computing network (Fig. \ref{fig:MHFL}). Enabling FL over such computing networks not only can augment AI experiences for urban mobile users, but also can democratize AI and make it accessible in a low-cost manner to everyone, including the large population of people in low-income communities, under-developed regions and disaster areas. 

\subsection{Challenges in Multi-hop Federated Learning}

Despite its great potential advantages to democratize AI,  \textit{multi-hop FL}, short for FL over multi-hop wireless edge computing networks, is still an unexploited area. The classic FL systems use single-hop wireless communications to directly connect to the edge servers or connect to edge routers that then reach the remote cloud servers via high-speed Internet core. In multi-hop FL networks, the end-to-end (E2E) model updates between the server and workers need to go through multiple noisy and bandwidth-limited wireless links. This results in much slower and nomadic model updates due to much longer and more random E2E delay. Such profound communication constraints fundamentally challenge the efficiency and effectiveness of classic FL systems as detailed below:

\begin{itemize}
\item \textbf{Degraded scalability of FL over wireless multi-hop networks}: The FL algorithms generally adopt a server-client architecture, where a central server collects and aggregates the model updates of all workers. The routing paths towards the central server can be easily saturated in wireless multi-hop networks due to the limited network bandwidth.In addition, different from classic distributed model training in data centers, FL exploits production networks to carry on model training traffic between workers and the server. Therefore, FL traffic has to compete with the background production network traffic (e.g., Internet traffic) for limited network bandwidth.  As a result, when the number of workers or the background traffic volume increases, network congestion will deteriorate progressively, which critically degrades the benefits of computation parallelization and slows down convergence speed.


\item \textbf{Difficulties of model-based optimization for multi-hop FL system}: Presently, there are limited research efforts on optimizing wireless FL systems. Existing efforts all focus on single-hop FL over cellular edge computing systems \cite{wirelessFL1,wirelessFL4,wirelessFL5}. With such assumption, the impact of wireless communication control parameters (e.g., transmission power) on the FL related metrics (e.g., model update delay and loss reduction) can be formulated in an explicit closed-form mathematical model, which greatly eases the FL system optimization. Such model-based optimization is not feasible in multi-hop FL, where the FL performance metrics (e.g., FL convergence time) cannot be explicitly formulated as a closed-form function of the networking control parameters, such as packet forwarding decision at each router.

\end{itemize}
 \begin{figure}
	\centering
	\includegraphics[width=2.5in]{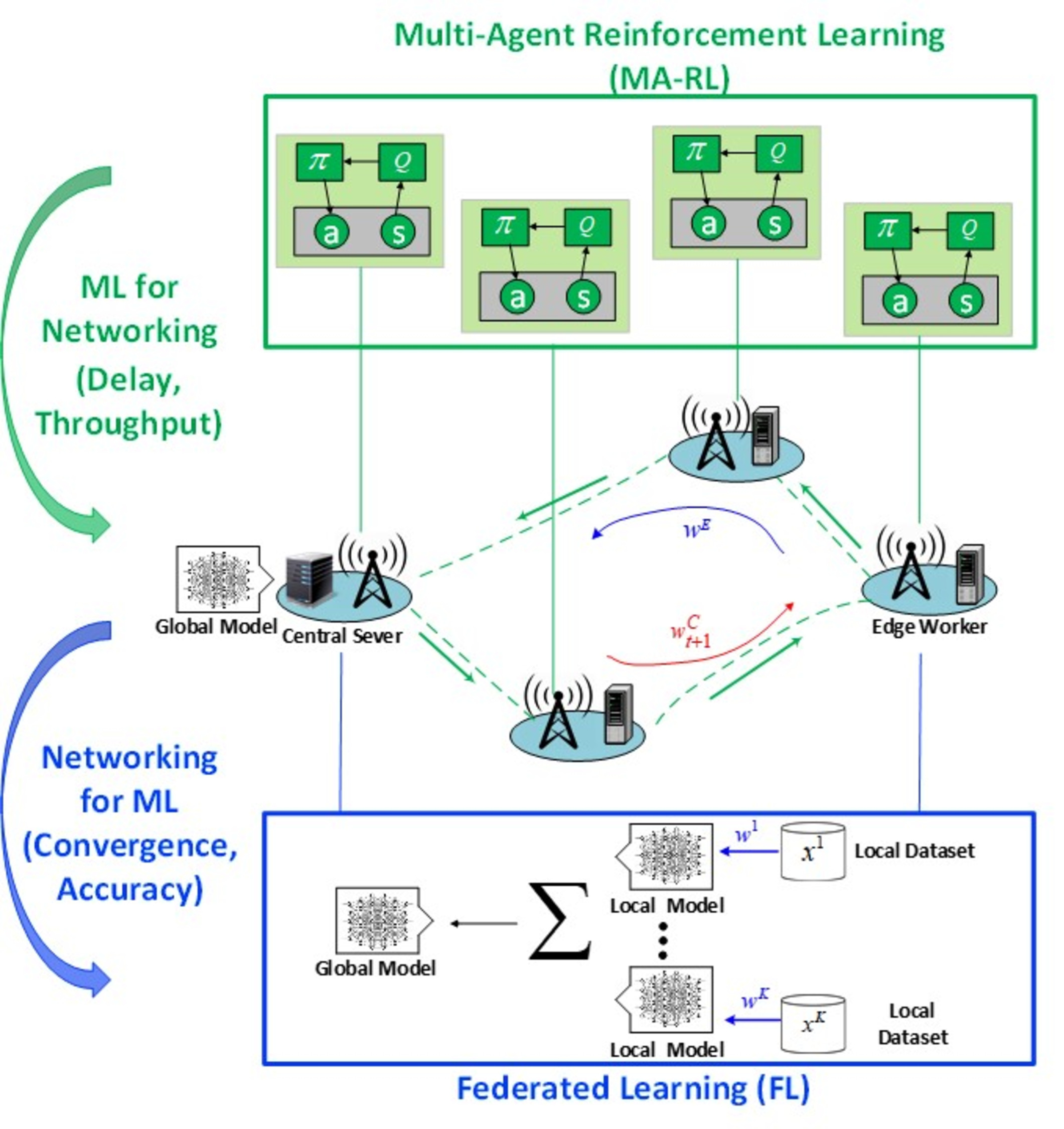}
	\caption{ Overall Architecture}
	\label{fig:WMN}
\end{figure}
\subsection{Our Contributions:}
The objective of this work is to develop a novel network-accelerated FL system over wireless edge by systematically and practically taming the networking-induced delay as shown in Figure~\ref{fig:WMN}.


\begin{itemize}

\item To the best of our knowledge, this is the first work in the literature to reveal, formulate, and experiment on the inherent interplay between multi-hop wireless networking and federated learning.
   
\item To minimize the FL convergence time, we exploit multi-agent reinforcement learning (MA-FL) for FL convergence optimization, which minimizes the networked-induced latency by learning the forwarding paths with the least delay for FL traffic flows. Moreover, our MA-FL agents are further optimized by action space refined via domain-specific knowledge for fast online RL training. 

\item We develop and prototype FedEdge, which is the first experimental framework in the literature for FL over multi-hop wireless edge computing networks. FedEdge thus enables  fast prototyping, deployment, and evaluation of novel FL algorithms along with machine learning-based FL system optimization methods in real-life wireless devices.

\item We implement the first physical experimental testbed in the literature for studying and testing FL over multi-hop wireless edge computing networks. The testbed is built on top of widely adopted Linux-based wireless routers and Nvidia computing nodes. Therefore, this testbed can provide valuable and broader insights into the practical performance of FL in the field.

\item We demonstrate via extensive experiments that the proposed FedEdge outperforms the FL system empowered by the SOTA production-grade wireless networking protocol and has the great potential to effectively improve the convergence performance of FL over wireless edge. 

\end{itemize}

The rest of our paper is organized as follows: Section~\ref{sec:algo} explains the runtime convergence of federated learning. Then, Section~\ref{sec:wfl} details our design and implementation of FedEdge framework, Section ~\ref{sec:eval} shows our system evaluation using routing application as a case study. Finally, Section~\ref{sec:conclusion} concludes our work. 

\section{Runtime convergence of Federated Learning} \label{sec:algo}
\subsection{Federated Learning via Regularized Local SGD}
\begin{figure}[ht!]
	\centering
	\includegraphics[width=1\linewidth]{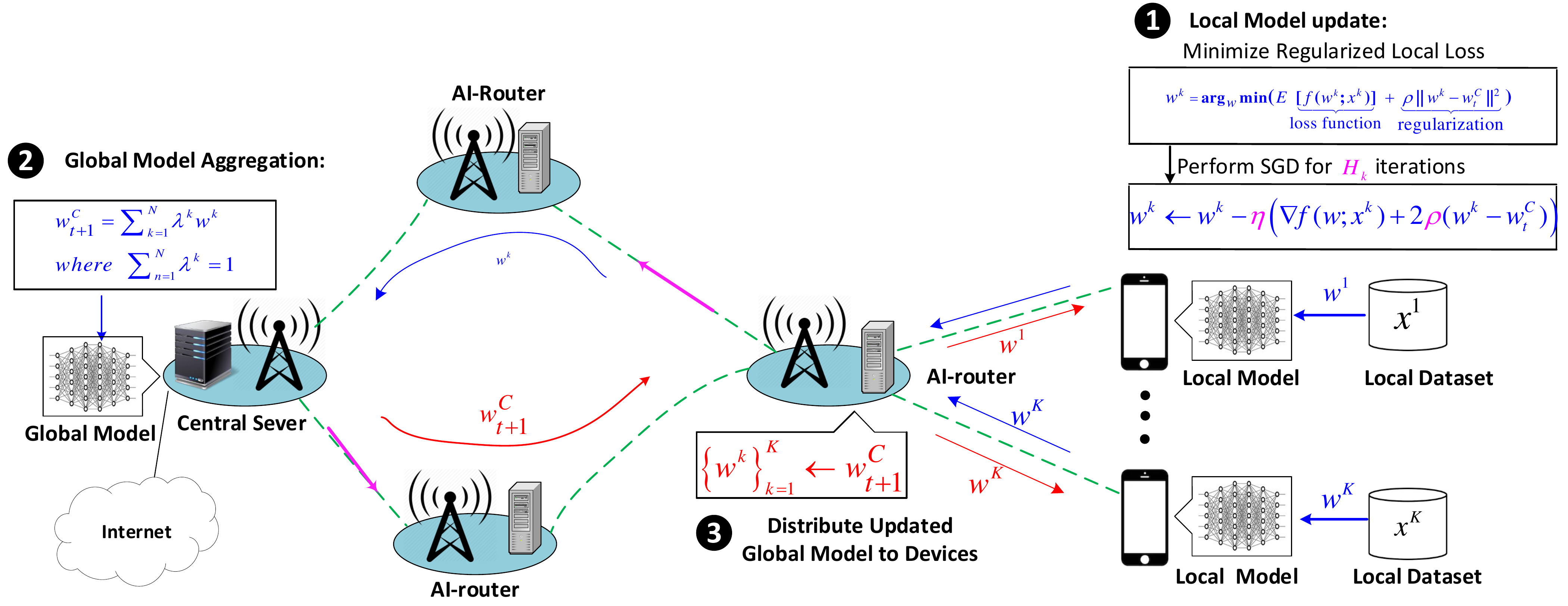}
	\caption{Federated learning via local SGD }
	\label{fig:FL}
\end{figure}
Federated learning methods are designed to handle distributed training of neural networks over multiple devices, where the devices have their local training data and aim to find a common model that yields the minimum training loss. Such a scenario can be modeled as the following distributed parallel
non-convex optimization
\begin{equation}\label{SGD}
\min_{w} F(w) = \sum_{k = 1}^{N}\lambda^kF^k(w), \ \ \ {F^k}(w) = {E}\left[ {f(w^k;{x^k})} \right]
\end{equation}
where $F(w)$ is the global loss, $F^k(w)$ is the local loss of device $k$, $N$ is the number of devices, $\lambda^k = \frac{n^k}{n}$ and $\sum_{k =1}^N\lambda^k=1$, where $n^k$ is the number of training samples on device $k$ and $n = \sum_{k} n^k$ is the total number of training samples in network. The local loss ${F^k}(w)$ is a non-convex function over data distribution ${x^k} \sim {\mathcal{D}^k}$, which is possibly different for different device $k$. The optimization problem in eq. \eqref{SGD} can be generalized by adding a quadric regularization term in the objective function \cite{AFL,fedprox}, i.e.,

\begin{equation}\label{RSGD}
 {F^k}(w) = E\underbrace {[f(w^k;{x^k})]}_{{\rm{loss}}} + \underbrace {\rho ||{w^k} - w_t^C|{|^2}}_{{\rm{regularization}}}
\end{equation}
where $w_t^C$ is the global model and $\rho $ is the penalty parameter that determines how much deviations from the global model the local model is allowed. 

To solve the above optimization problem, FL methods follow a common stochastic optimization technique, called {local SGD}, which alternates between {local SGD iterating} and {global model averaging} for multiple {(server-worker communication) rounds}, where the worker is the device that participates in the collaborative model training. As shown in Figure~\ref{fig:FL}, during each round, the worker tries to reduce its local loss $F^k(w)$  by performing $H_k$ mini-batch SGD iterations with each iteration updating the model weights, following: 

\text{Local SGD Iterating:} 

\begin{equation}\label{local_it}
{w^k} \leftarrow {w^k} - \eta \frac{1}{B}\sum\nolimits_{x^k \in \mathcal{I}^k}\left( {\nabla f(w^k;{x^k}) + 2\rho ({w^k} - w_t^C)} \right)
\end{equation}
where $\mathcal{I}^k$ is a subset (mini-batch) of the training samples on worker $k$ and $B = |\mathcal{I}^k|$ is the size of the mini-batch. After finishing $H_k$ local SGD iterations, the workers send their local models $\{w^k\}_{k \leq K}$ to the central server, which averages them and updates the global model accordingly
\begin{equation}\label{global_it}
\text{Global Model Averaging:\ \ } w^c = \sum_{k = 1}^{K}\lambda^kw^k 
\end{equation}
where $K$ is the number of devices selected to be the workers. The new global model is sent to the workers and the above procedure is repeated. 

It is worthy to note that minimizing regularized loss ensures that the local workers will not fall into the model update trajectories that are far away from the current global model. Such practice can effectively prevent the potential divergence caused by statistical heterogeneity \cite{AFL} and system heterogeneity \cite{fedprox}. On the one hand, the workers involved in FL training tend to possess significantly diverse data samples so that they follow unbalanced and non-IID data distribution, thereby introducing statistical heterogeneity. On the other hand, the workers generally possess diverse computation resources (e.g., CPU, GPU and RAM). To mitigate the blocking effects of stragglers (slow workers) and reduce computation-induced latency, each worker can perform different number of local iterations $H_k$ according to its computation constraint, which leads to system heterogeneity. When the penalty parameter equals to zero, i.e., $\rho = 0$ and all workers adopts the uniform local updates, i.e.,  $H_k = H, \ \forall k \leq N$, then the local SGD algorithm in e.q., \eqref{local_it} and \eqref{global_it} becomes the classic FedAvg algorithm \cite{FL}.

\subsection{Convergence of Local SGD}

\subsubsection{Iteration Convergence} Before local SGD is applied  in FL settings, it  already showed very promising performances for distributed optimization in data center environments. The key advantage of local SGD is {its low communication overhead} along with {high convergence speed}. Recent research shows that for non-convex optimization with both IID and non-IID data, local SGD can achieve fast \textbf{$\mathcal{O}(1/\sqrt{KT})$} convergence \cite{convergenceFL1,convergenceFL2}, i.e., achieving linear speedup w.r.t. the number of workers $K$, where $T$ is the total number of iterations performed by each worker. This is the optimal convergence performance achieved by the celebrated {parallel mini-batch SGD} methods \cite{pSGD1,pSGD2}, where each worker sends its model or gradient to the server after each local SGD iteration is done.  Therefore, parallel mini-batch SGD achieves the optimal \textbf{$\mathcal{O}(1/\sqrt{KT})$}  convergence at the cost of $T$ communication rounds. However, to achieve the same convergence performance, local SGD only needs  \textbf{$\mathcal{O}(T^{3/4}K^{3/4})$} communication rounds \cite{convergenceFL1,convergenceFL2}. In other words, local SGD can preserve the fast convergence with significant less communication cost by putting more computation loads on the workers, i.e., by letting workers perform \textbf{$\mathcal{O}(T^{1/4}/K^{3/4})$} local SGD iterations instead of one.

\begin{figure*}[t]
\captionsetup{singlelinecheck = false, justification=justified}
  	\centering
      \vspace{-0.3cm}
  	\subfigure[]{%
  	\label{fig:high_inc_1}%
  	  \includegraphics[width=0.25\linewidth]{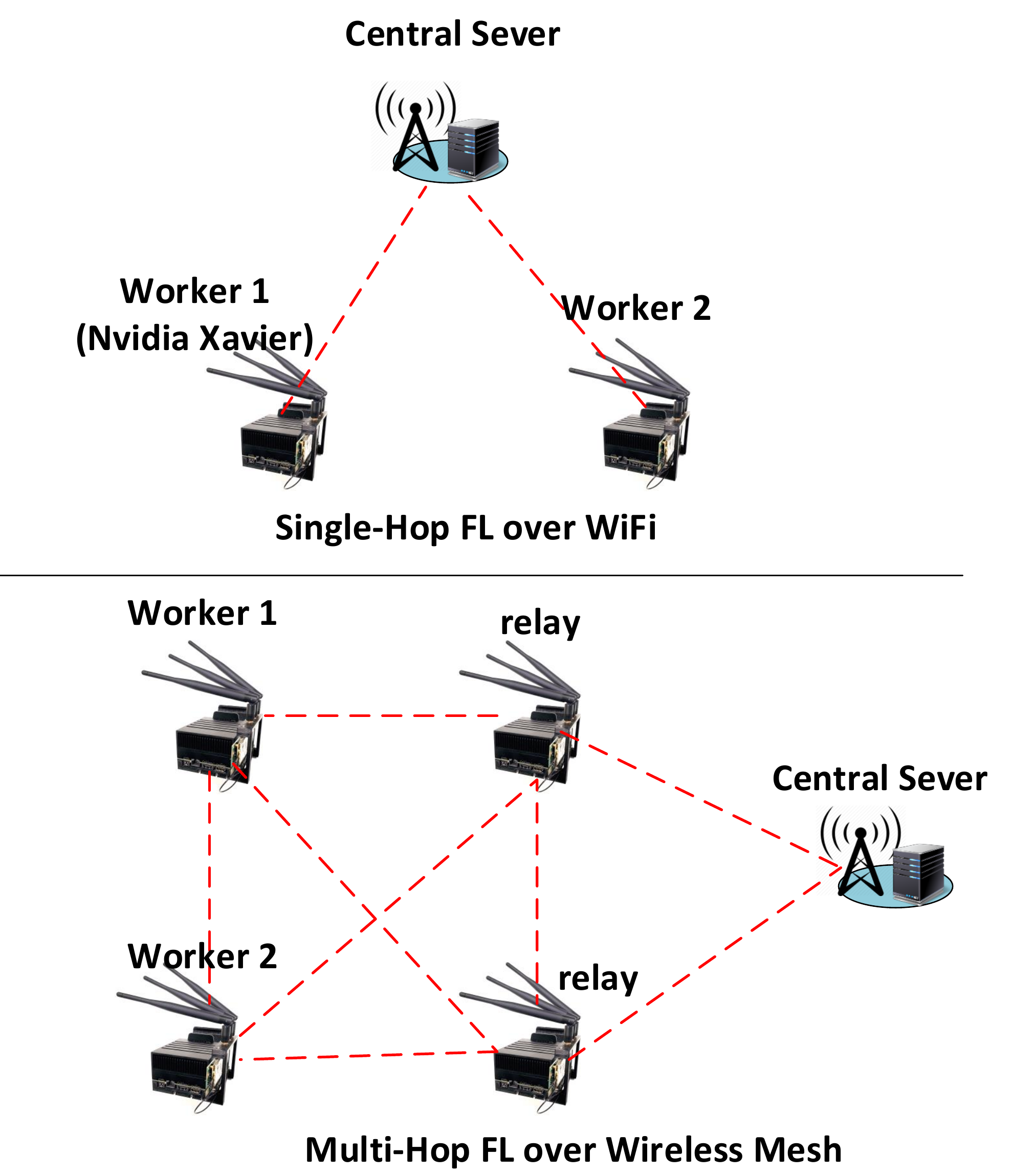}
  	}
  	\vspace{-0.3cm}
  	\subfigure[]{%
 	 \label{fig:high_inc}%
 	\includegraphics[width=0.3\linewidth]{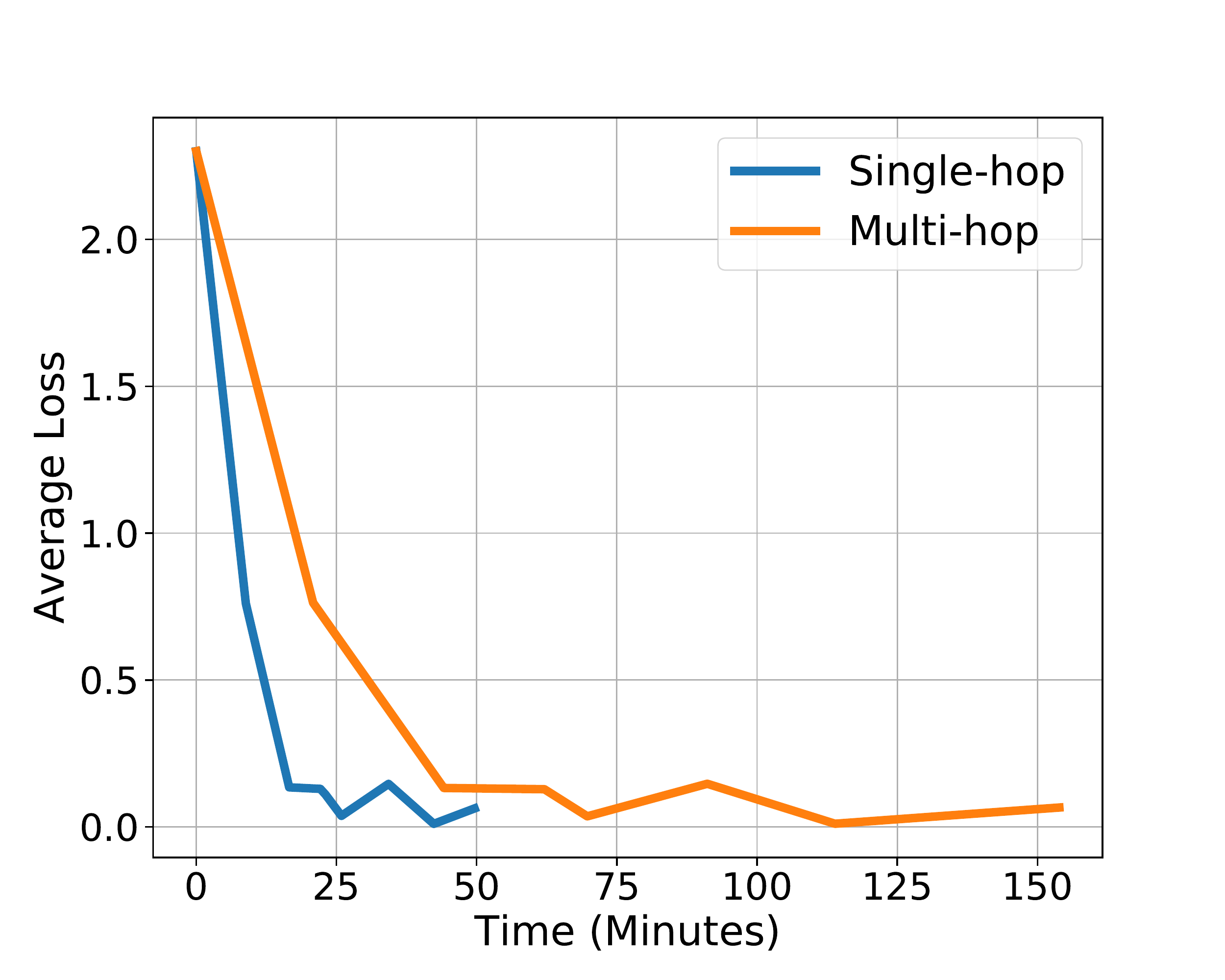}
  	}
  	\vspace{-0.3cm}
  	\subfigure[]{%
 	 \label{fig:high_inc}%
 	\includegraphics[width=0.3\linewidth]{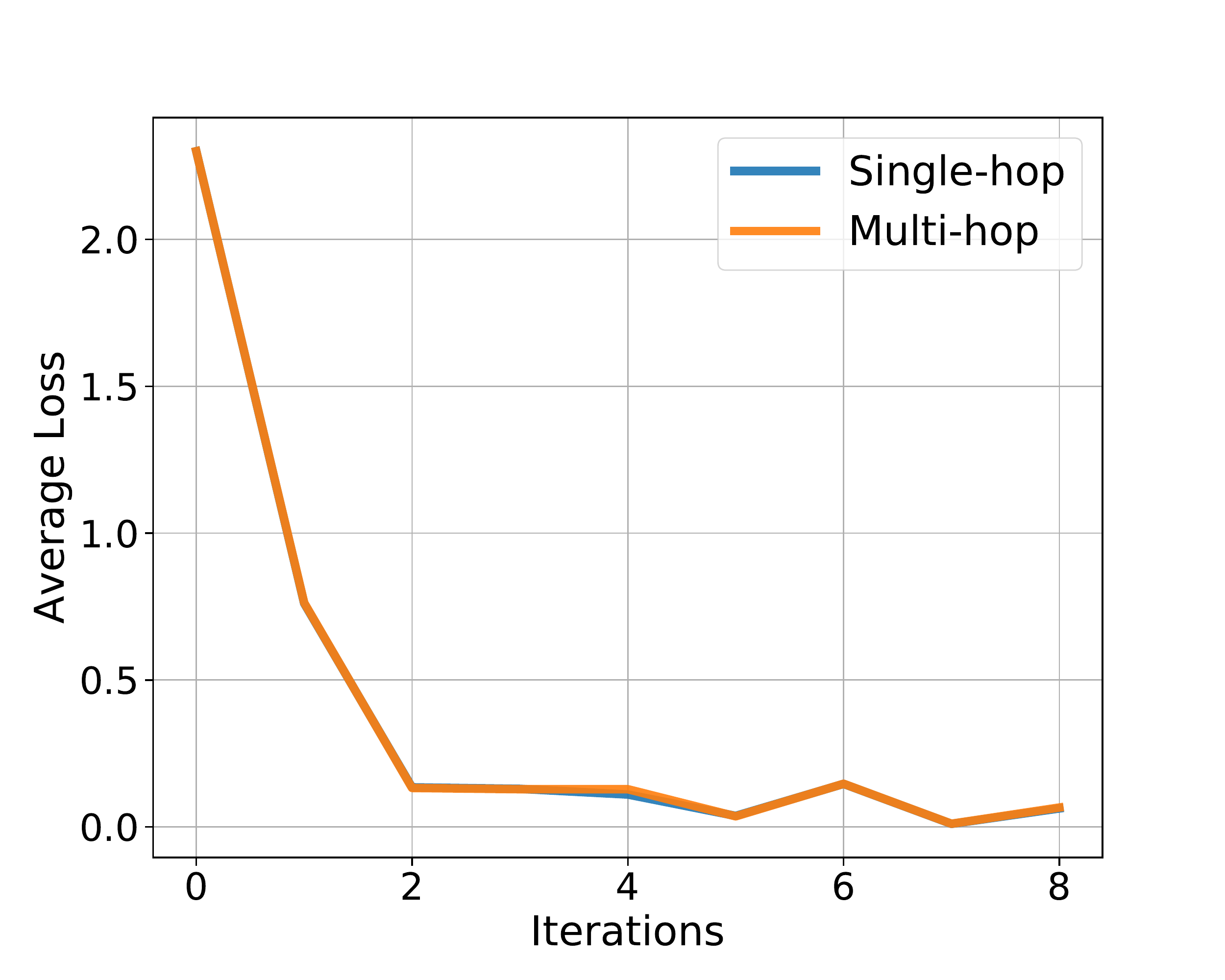}
  	}
  	\vspace{3mm}
  	\caption{We use Nvidia Xavier nodes to implement two workers and one server to train MNIST dataset for digit recognition task. To test the impact of wireless networking, we use the exactly same parameters (e.g., initial model weights, number of rounds, number of local iterations, batch size, and learning rate) for FL over single-hop and multi-hop wireless networks, respectively. (\textbf{a. Network topology}) for single-hop networks, we use IEEE 802.11ac for wireless connections. For multi-hop mesh networks, we use IEEE 802.11s for multi-hop routing, which still uses IEEE 802.11ac for MAC/PHY functions. The testbeds are deployed in the first floor of the UNCC CS department with interferences from co-existing campus WiFi networks (\textbf{b. Runtime Convergence}) The wireless multi-hop FL converges much slower with respect to the true training runtime (wallclock time)  (\textbf{c. Iteration Convergence}) The single-hop and multi-hop FL systems have the same iteration convergence performance.
  	}
  	\label{fig:high_inc_all}
  \end{figure*}

\subsubsection{Runtime Convergence of Local SGD}
 Local SGD method (e.g., de-factor FL algorithm FedAvg) is generally implemented in a synchronous manner, where the SGD update sequences on the workers are synchronized (by model averaging). In other words,  the server needs to wait for the model updates from all workers and then it can perform model aggregation, after which the workers can resume their local SGD updates for the next round. As a result, if the actual training runtime (wallclock time) $t$ is used instead of iteration index $T$,  the convergence of local SGD could be as poor as \textbf{$\mathcal{O}(\sqrt{\tau_{max}}/\sqrt{Kt})$}(where each worker only performs one local iteration). $\tau_{max}$ is the delay of the slowest worker (straggler) to deliver its local model to the server, which could be very small in high-speed data center networks and wireless single-hop networks (e.g., WiFi or cellular). In wireless multi-hop networks, $\tau_{max}$  becomes a more dominant factor affecting the true runtime convergence due to the large, random and heterogeneous E2E communication delays experienced by the workers.  As a result, {the theoretically fast convergence of local SGD can be practically slowed down in wireless multi-hop networks}. Such projection is also verified through a simple experiment (Fig.~\ref{fig:high_inc_all}). Moreover, the linear convergence speedup by increasing the number of workers $K$  could also be accompanied by the increased delay $\tau_{max}$ due to escalated network congestion, which leads to  convergence slowdown.

\section{Optimizing FL Convergence via Reinforcement Learning}
\subsection{Problem Formulation}
Our overall objective is to minimize the run-time convergence time to achieve the desired FL accuracy. Towards this goal, the optimal strategy is to minimize the worker-server delay of the slowest worker, which experiences the maximum delay among all workers. However, in  highly dynamic wireless environments, the role of the slowest one can be randomly switched among different workers as time proceeds. In this paper, we sought a sub-optimal solution, where we minimize the average end-to-end (E2E) delay between all workers and the server. However, even for such sub-optimal solution, we cannot apply the classic model-based optimization because the server-worker E2E delay cannot be explicitly formulated as a closed-form function of the routing/forwarding decisions. As a result, a model-free optimization strategy based on multi-agent reinforcement learning is much more desirable, where each wireless router exploits its instantaneous local experiences to collaboratively learn the delay-minimum routing paths between the workers and the server.

In particular, this problem can be formulated as the multi-agent Markov decision processes (MA-MDP), which can be solved by multi-agent reinforcement learning algorithms. Given the local \textit{observation} $o_i$, which is the source IP and destination IP of the incoming FL packet, each router or \textit{agent} $i$  selects an \textit{action} $a$, i.e., the next-hop router, to forward this packet, according to a local forwarding \textit{policy} $\pi_i$. After this packet is forwarded, the router $i$ receives a \textit{reward} $r_i$, which is the negative one-hop \textit{delay} between router $i$ and the selected next-hop router. The packet delivery delay $d_{i,i + 1}$ is the time interval between the time when packet arrives at router $i$ and the time when the packet arrives at the next-hop router $i + 1$. The packet delivery delay $d_{i,i + 1}$, which includes the queuing delay, processing delay and transmission delay, is a random value measured in real-time by in-network telemetry module introduced in the next section.  The \textit{return} $G_{i} = \sum_{k = i}^{T}r_k$ is the total reward from intermediate state $s_i$ to final state $s_T$, where $s_i$ and $s_T$ are the states when a FL packet arrives at the relay router $i$ and destination router $T$, respectively. Let $s_1$ be the initial state when a FL packet enters the network from its source router. The source/destination router is the router that a worker or the server is attached to.
The \textit{objective} is to find the optimal policy ${\pi}_i$ for router $i$ so that the expected return $J(\bm{\pi})$ from the initial state (i.e.,E2E server-worker delay) is optimal, where
$J(\bm{\pi}) = E[G_1|\bm{\pi}] = E[\sum\nolimits_{i = 1}^{T}r_i |\bm{\pi}]$
where $\bm{\pi} = \pi_1,...,\pi_N$. 
\subsection{Convergence Optimization via Multi-agent Reinforcement Learning}
To solve the above MA-MDP problem, we exploit the multi-agent reinforcement learning, where the routers (agents) distributively learn the optimal target forwarding policy $\bm{\pi}$ to minimize the average server-worker delay. To implement the multi-agent reinforcement learning algorithm, we adopt a distributed actor-critic architecture similar to asynchronous advantage actor-critic
(A3C) \cite{champ_NI19,mnih2016asynchronous}, where each router individually runs a local critic and a local actor,


\subsubsection*{Local Critic for Policy Evaluation}
 The performance of the policy $\pi$ is measured by the action-value $q_i^{\pi}(s,a)$, which is an E2E TE metric. The action-value $q_i^{\pi}(s,a)$ of router $i$ can be written as the sum of 1-hop reward of router $i$ and the action-value of the next-hop router $i+1$, i.e.,
\begin{equation}\label{bellman}
q_i^{\pi_i}(s,a)  = E\left[r_i + q_{i+1}^{\pi_{i+1}}(s', a')\right].
\end{equation}
By applying exponential weighted average, the estimate of  $q_i^{\pi_i}(s,a)$, denoted by $Q_i^{\pi_i}(s,a)$, can be  updated based on 1-hop experience tuples $(s,a, r_i,s',a')$ and the estimate of $q_{i+1}^{\pi_{i+1}}(s', a')$ of next-hop router, denoted by  $Q_{i +1}^{\pi_{i + 1}}(s',a')$, i.e.,
\begin{equation}\label{EWA}
\ Q_i^{\pi_i}(s,a) \gets Q_i^{\pi_i}(s,a) + \alpha [r_i + Q_{i +1}^{\pi_{i + 1}}(s', a') - Q_i^{\pi_i}(s, a)] 
\end{equation}
where $\alpha \in (0 ,1]$ is the learning rate.

\subsubsection*{Local Actor for Policy Improvement}
Based on critic’s inputs, the local actor improves the local policy, which aims to maximize the cumulative sum of reward $J(\bm{\pi})$. This can be done by applying greedy policy, where  each router $i$ greedily improve its current policy $\pi_i$, i.e., select the action with the maximum estimated action-value,
\begin{equation*} 
\pi_i(s) \gets \arg\max_{a}Q_i^{\pi_i}(s,a).
\end{equation*}
Besides greedy policy, we also exploit two near-greedy policies to encourage exploration. The first one is $\epsilon-$greedy policy with exponential decay. With such policy, the router selects the greedy action defined in eq. \eqref{EWA} with probability $1 - \epsilon(t)$ and select other actions with probability $\epsilon(t)$. The $\epsilon$ decays exponentially as time proceeds, i.e., $\epsilon(t) = \epsilon_0\beta^{t}$, where $0 < \epsilon_0 < 1$ and  $0 < \beta < 1$. The second near-greedy policy is softmax-greedy policy, where each action $a$ is selected with a probability $P(a)$ according to the exponential Boltzmann distribution,
\begin{equation}\label{softmax}
P(a) = \frac{\exp(Q_i^{\pi_i}(s,a)/\tau)}{\sum_{b \in \mathcal{A}_i}\exp(Q_i^{\pi_i}(s,b)/\tau)}.
\end{equation}  
where $\tau$ is the temperature. The actions performed by the router are generated  according to the behavior policy, which is either same as the target policy $\pi_i$ or similar to the target policy but more exploratory (on-policy learning). For off-policy learning, the target policy is generally the greedy policy and the behavior policy is generally near-greedy to enable explorations. 

\begin{figure} 
\captionsetup{singlelinecheck = false, justification=justified}

	\centering
	\includegraphics[width=1\linewidth]{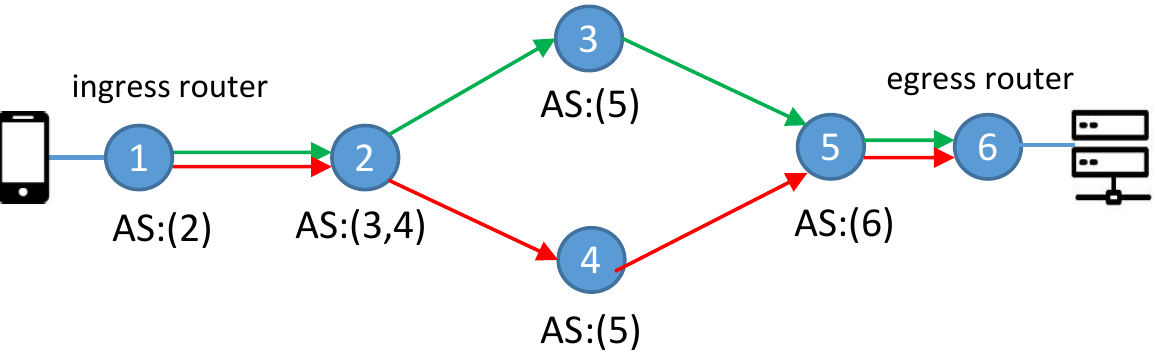}
	\caption{Loop-free action space (AS) refining. There exist two loop-free paths between the ingress router and egress router. For each router, we refine the action space (AS) for each router in the network that two routing paths traverse through.The RL algorithm will explore the actions in the refined action space to learn loop-free routing paths.}
	\label{fig:action_space}
\end{figure}

\subsection{Loop-free Action Space Refining}
The MA-RL routing is one kind of the distributed routing algorithms. However, the key idea of RL is to improve the policy by learning from experiences including failures. Therefore, an agent can freely explore and learn all possible routing paths including the ones with loops, where the data packets continue to be routed within the network in an  circle. Our experiments show that the routing loops can have a catastrophic impact on a RL-based networking, such as the slowly converged routing policy and TCP disconnections between the server and workers. To address this problem, we propose a action space refining algorithm, which aims to construct the loop-free action spaces for each router in such a way that the routers can independently and distributively explore any forwarding action (i.e., the next-hop router) from such refined action space, while avoiding generating routing paths with loops. The refined action space is defined with respect to each pair of ingress and egress routers. The ingress router is the router from which the FL traffic flow enters the network and the egress router is the router from which the FL traffic leaves the network. Therefore, the maximum number of action spaces constructed on each router is equal to $2N$, where N is the total number of routers in the network. The action space refining algorithm works as shown in the Fig. \ref{fig:action_space}. First, build the global network topology. Then, find all the loop-free paths between the ingress router and egress router by applying iterative depth-search-first (DSF) traversal or K-shortest path finding algorithms (with sufficiently large K). Next, for each router, there may exist multiple paths traversing it and its action space is a set of the next-hop nodes of all the traversing paths. It is easy to prove that employing such refined and loop-free action spaces, our MA-RL forwarding scheme will surely learn the routing paths without loops. 


It is worth to note that the action space refining algorithm is only performed by a network controller which has the global network topology. The implementation details are shown in the next section.

\section{FedEdge Design and Prototyping}\label{sec:wfl}

\subsection{FedEdge Overall Design}

Existing FL experimental frameworks (e.g., TensorFlow Federated (TFF)  \cite{tff}, PySyft\cite{pysyft}, LEAF \cite{leaf}, and FedML \cite{he2020fedml}) only support experiments of federated learning without taking into account the impacts of wireless federated networking. What is more important, they did not support the reconfiguration of the network stack and did not allow us to incorporate new networking schemes. Therefore, they are not suitable to study the interplay between federated networking and federated computing. In addition, the frameworks mentioned above were highly dependent on WebSockets for their communication, which leads to a reactive disconnection between the worker and server when their links experience a short-lived delay. Moreover, except FedML, these frameworks can not be readily deployed on the physical edge devices without cumbersome modifications.  To address the above challenges, we develop and prototype FedEdge, which is the first experimental framework for wireless multi-hop FL.
The full pipeline of prototyping, deployment, and evaluation is expected to be easily performed  with real wireless devices given the FedEdge framework.

\begin{figure} 
	\centering
	\includegraphics[width=1\linewidth]{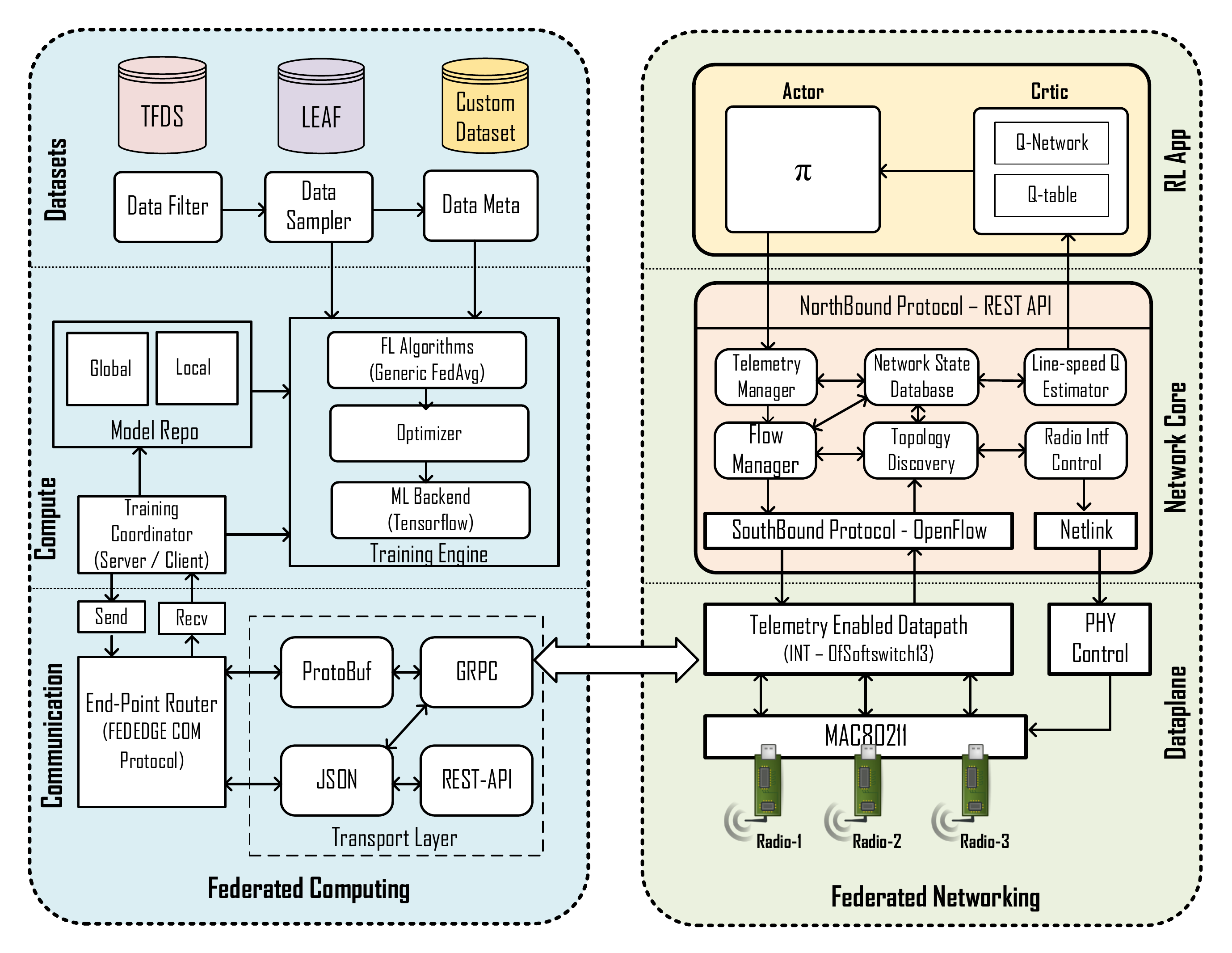}
	\caption{Architecture of FedEdge Framework}
	\label{fig:FedEdge}
\end{figure}
 
FedEdge has two unique features. The first feature is its modularity for both communication and computation functions. This facilitates the simultaneous evolutions of federating computing and federated networking solutions and allows us to evaluate the complex FL system under diverse network and computation conditions. Second, FedEdge is a highly programmable experimentation platform that can be easily deployed on real wireless systems. With these two  features, we can design and deploy new FL algorithms along with customized datasets, preprocessing schemes, and training pipelines. Simultaneously, we can  innovate on various networking mechanisms and test their impact on the performance of federated learning systems. FedEdge will be open-sourced after we make it well documented.

As shown in Figure \ref{fig:FedEdge}, FedEdge consists of two key components: federated computing, which customizes and configures FL-related functions (e.g., FL training algorithm setup including node role selection, model selection, and dataset preparation) and federated networking, which is a distributed AI-oriented wireless network operating system. Federated networking is responsible for providing fast wireless networking connections between the aggregator and workers. What is more important, federated networking system is designed in nature to facilitate AI-empowered networking optimization, including customizable actor-critic RL agent for instantiating a variety of AI-enabled routing algorithms, in-band telemetry that enables cost-efficient data collections for online RL training, and programmable routing table (datapath) for real-time RL policy executions.

Each FedEdge component follows a layered design. In particular, FL engine consists of three layers. (1) FL Datasets layer stores the datasets for federated training (2) FL compute layer provides vital functions to train the model and save the models and (3) FL communicate layer establishes logical and reliable connections between workers and aggregator by using the proposed FedEdge communication protocol (FedEdge COMM). Federated Networking design is composed of 3 layers. (1) Dataplane incorporates SDN software switch with our proposed in-band telemetry scheme to enable programmable packet forwarding, while simultaneously providing low-cost real-time collection and reporting of network state measurements. (2)  Network Core provides essential networking services and functions. such as topology discovery, network state database, and traffic flow management. (3) RL application layer hosts the actor-critic RL agent to enable delay-optimal routing between the aggregator and the workers.  

\subsection{Federated Computing}
\subsubsection{\textbf{FL Datasets}}
Federated learning is based on the well-established machine learning technique with a key significant difference in how the data is used to train the models. One of the FL technique’s primary objectives is to protect the user’s privacy by avoiding sharing data from the origin device.  Firstly, without reinventing the wheels, we leverage the existing datasets hosted by Tensorflow TFDS \cite{tfds} and LEAF \cite{leaf}, both of which are well known in the FL community. We also provide extended APIs to incorporate custom datasets specific to the experimenter to integrate within our FL framework.

\subsubsection*{Dataset-Setup}
In our FL training pipeline, the first step is to distribute the datasets to different workers in the training process. To do so, the experimenter has to specify three primary parameters, which are (1) Number of workers involved and (2) the Data distribution type - I.I.D or non-I.I.D and (3) Dataset name and the repository of the dataset (TFDS, LEAF,  or CUSTOM). Next, our FedEdge Datasets module will split the dataset respectively such that data is readily available for training. Data for the worker is selected at the beginning of the FL training procedure. In this data staging process, the user can further preprocess the data using FedEdge's builtin functions to encode, normalize, and standardize the data.  Such synthesized data is then gathered into batches by the batch size specified during the initial training epoch.  Furthermore,  batches are converted to a Tensorflow-accelerated data pipeline to avoid the IO bottleneck,  thereby accelerating the runtime of each epoch.

\subsubsection*{Pipeline}
In the FL training process, the experimenter can choose the data for training in an arbitrary fashion for each global round. For each global round, the aggregator may specify the class and number of samples used for the current epoch. With such flexibility in the training process, we provide a highly customizable pipeline for consuming data during the training process with two sub-modules including data filtering and sampling. At each round of global training step, the aggregator can pass parameters such as the data class to filter and sampling technique for the training process. In this case, the sampling procedure will determine whether the entire filtered data should be used or have to be sub-sampled to limit the number of samples consumed for the training step. To simplify the above-mentioned process, we consider all datasets will contain a META which provides the key statistics about the dataset, such as the number of classes and the number of samples for each class and in total.

\subsubsection{\textbf{FL Compute}}

Our motivation for designing FedEdge compute module is to provide an extensible processing stack such that it is easy to customize the training pipeline without being limited to the communication protocol. Compute module is comprised of 3 stages of processing: (1) Training Coordinator, (2) Model Repo, and (3) Training engine. In the following section, we will briefly describe the functionality of each stage.

\subsubsection*{Training Coordinator}
Federated learning involves nodes with two types of roles, server/aggregator and worker. The key functionality of  training coordinator is to set up the Federated computing framework based on the role such that the worker nodes execute model training and server nodes perform only model aggregation and model evaluation. In addition, it also handles the complete training cycle by setting up the training engine and storing and retrieving models from the Model Repo. 

\subsubsection*{Model Repo}
Federated training procedure mandates frequently exchanging model under training between worker and the aggregator. Some FL algorithms require the model of current round to be updated using the models from the previous training rounds. To facilitate such procedure, we implemented a Model Repo that stores the global and local models for a specified time duration. Each model is time-stamped before writing to the Repo so that it is easy to distinguish the current model and historical models.

\begin{algorithm}
	\caption{Local SGD - Aggregator}
	\begin{algorithmic}[1]
		
		\renewcommand{\algorithmicrequire}{\textbf{Input:}}
		\renewcommand{\algorithmicensure}{\textbf{Output:}}
		
		\REQUIRE maxround \textit{{r\textsubscript{m}}}, worker epoch \textit{H\textsubscript{k}, k $\leq$ N }, batch size \textit{B}
		\ENSURE Global Model  $W$\textsuperscript{c}
		
		\textbf{AGGREGATOR PROCESS}
		\STATE{Initialize Worker Registry: $R$}
		\STATE {Initialize Current Model Queue: $Q_{fresh}$ }
		\STATE {Initialize Worker State Queue: $Q_s$}

		\STATE \textit{WORKER NODE REGISTRATION}
		
		\FOR {round $r \leq  r\textsubscript{m}$ }
		\IF {r = 1}
		\STATE \textit{Initialize} : w\textsuperscript{c}
		
		\FOR {\textit{k} in $R$ \textbf{in parallel}}
		\STATE $updateWorker \gets w\textsuperscript{c}$
		\STATE $Q\textsubscript{s} \gets GLOBAL\_MODEL\_RECV(k)$
		\ENDFOR
		
		
		\ELSE    
		\STATE \textit{Wait local models from workers}
		\FOR {\textit{k} in $R$ \textbf{in parallel}}
		\STATE \textit{Ask workers to start training} : 
		\STATE $\textit{Q}_{fresh} \gets \textit{w}^k \gets \textit{train}(\textit{k},\textit{H}\textsubscript{t}, \textit{B})$
		\STATE \textit{local model received from worker} : 
		\STATE $Q\textsubscript{s} \gets LOCAL\_MODEL\_RECV(k)$
		\ENDFOR
		\STATE \textit{Perform Model Aggregation}
		\STATE $w^c_{t}   \gets \sum_{k = 1}^{R}\lambda^kw^k$ 
		\STATE \textit{Send updated global model to workers} : 
		\FOR {\textit{k} in $R$ \textbf{in parallel}}
		\STATE $updateWorker \gets w^c $
		\STATE $Q\textsubscript{s} \gets GLOBAL\_MODEL\_RECV(k)$
		\ENDFOR
		\ENDIF
		\ENDFOR
		
		\RETURN \textit{W}\textsuperscript{c}
	\end{algorithmic} 
	\label{algo:aggregator}
\end{algorithm}

\begin{algorithm}
	\caption{Local SGD - Worker}
	\begin{algorithmic}[3]
		
		\renewcommand{\algorithmicrequire}{\textbf{Input:}}
		\renewcommand{\algorithmicensure}{\textbf{Output:}}

		\REQUIRE workerid \textit{{k}}, worker epoch \textit{H\textsubscript{k}}, batch size \textit{B}
		\ENSURE Local Model  $w$\textsuperscript{k}

		\textit{Initialize Status Queue: $Q_w$}
		\\ \textit{Initialize Dataset Store: $D_S$}
		\STATE \textsf{REGISTER}(workerid)
		\STATE \textbf{FUNCTION} $\textit{train}(\textit{k}, \textit{H}\textsubscript{k}, \textit{B})$
		\STATE $Q_w \gets TRAINING\_STARTED$
		\WHILE {i $<$ \textit{H\textsubscript{k}}}
		\FOR {\textit{bs} in D\textsubscript{s}}
		\STATE ${w^k} \leftarrow {w^k} - \eta \frac{1}{B}\sum\nolimits_{x^k \in \mathcal{I}^k}\left( {\nabla f(w^k;{x^k}) + 2\rho ({w^k} - w_t^C)} \right)$
		\ENDFOR
		\ENDWHILE
		
		\STATE $Q_w \gets TRAINING\_FINISHED$
		\RETURN $w^k$ to AGGREGATOR
	\end{algorithmic} 
	\label{algo:worker}
\end{algorithm}

\subsubsection*{Training Engine}  The operations of the Training Engine will vary based on the role of the node. To further simplify the context of FedEdge Compute, we describe the functions of FedEdge Compute based on the node’s role in the training process, and the sequence of communication among the nodes are shown as the sequential flow in Figure \ref{fig:comapi}. The detailed operations of FL training are shown in Algorithm \ref{algo:aggregator} and \ref{algo:worker}.

FedEdge aggregator node is the central node that controls the life cycle of FL training. The aggregator has a crucial role in initiating, coordinating, and monitoring the FL training cycle. Before beginning the FL training cycle, each worker should register their IDs, a combination of IP and port numbers to the aggregator's worker registry built into the FedEdge Communicate End-point router module. This registry is constructed using a hash map that stores worker ID as the key and their communication HTTP/GRPC end-point as the value. Only registered workers can participate in the training cycle. The aggregator performs a two-stage process to launch the FL training sequence: (1) Construct the model and upload it to the workers (2) Launch the training with the user-supplied training configuration. In the following section, we briefly discuss the two-stage process:

\begin{itemize}

\item \textit{Stage-1:} Our FedEdge compute module provides in-built models for image classification tasks using convolutional neural network (CNN). Besides, users can easily modify the in-built models and integrate their customized loss functions.  Each model will then be trained with the user's choice of optimization algorithms. Currently, FedEdge compute module provides in-built support for the regularized local SGD algorithm by default, which can be considered as the generalized FedAvg. This newly created model is then shared with the workers for training. To send the model, the aggregator revisits the worker registry to obtain each worker’s HTTP/GRPC end-points. If the worker successfully received the model, then the corresponding worker state will be updated based on the message from the worker with the context $GLOBAL\_MODEL\_RECV$.

\item \textit{Stage-2:}
To begin the FL training cycle, user needs to supply the following parameters at the minimum: global rounds,  local rounds per node, dataset repo and dataset,  model to train, and data partition type. The global round controls the total number of rounds workers will use to train the shared model, and local rounds define the number of epochs each worker will use to update the local model. Since our FedEdge framework hosts datasets from a variety of repo's, users should specifically mention the dataset and the repo used for the experiment. With the user-supplied parameters, a training cycle is constructed where a cycle is defined as (1) launching a global round (2) wait to receive models from all workers and  updates the state of the corresponding worker to $LOCAL\_MODEL\_RECV$. (3) perform model aggregation, and (4) finally share the aggregated model with the workers. This cycle will terminate once the required number of global rounds are reached or when the target accuracy is achieved.

\end{itemize}


FedEdge worker node operates on the request of the aggregator node. FedEdge aggregator interacts with the worker node to initiate the training sequence by sharing the global model and initial training parameters. The model received from the aggregator node will be stored as a global model into the model repo. Worker node loads the global model and creates its copy of the local model by cloning. This local model is then used to train over multiple epochs with the supplied optimization algorithm to minimize the local loss function. Finally, the updated model is shared with the aggregator.

\subsubsection{\textbf{FL Communication}}
The reliability of the FL system is heavily dependent on the robust communication stack for collaborative learning. As more research efforts are focused on improving the overall FL system performance by improvising computation, little to no effort has been targeted towards optimizing the communication layer. We believe the lack of a flexible platform is one of the inherent challenges to pursue research in this venue. Extendability and ease of programmability as the objective, we adopted modular design which consists (1) Send and Recv  (2) End-point router and (3) FL-Transport Layer as the submodules. In the following section, we will briefly discuss the functionality of the modules as mentioned earlier:

\subsubsection*{Send and Recv}
FL communication module interacts with other modules within the federated computing framework using send and recv submodules. The key objective of these submodules is to serve as a interface between communication  and compute layer. While the operation of this submodule is much simpler for the worker's role, it is quite complex for the aggregator's role as it requires non-blocking communication sessions for scalability. To realize non-blocking and parallel communication sessions, we developed this submodule using asynchronous python library ASYNCIO\cite{asyncio} and FastAPI\cite{fastapi}. 
 \begin{figure}[h!] 
	\centering
	\includegraphics[width=0.8\linewidth]{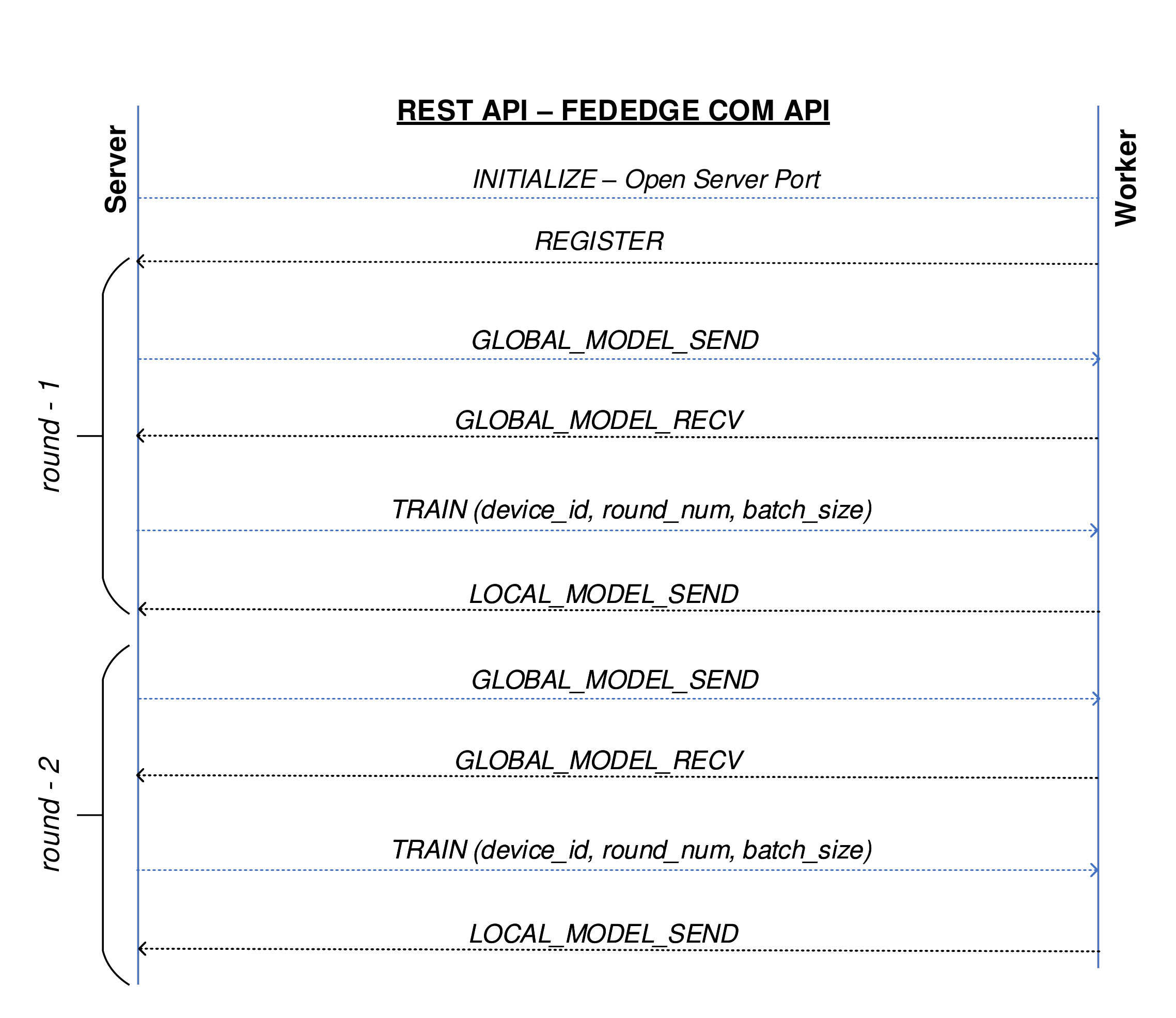}
	\caption{FedEdge Communication Protocol}
	\label{fig:comapi}
\end{figure}
\subsubsection*{End-point router}
FedEdge platform exposes functions for FedEdge Compute and dataset layers through an End-Point router (EPR). This routing layer routes messages from external nodes to the respective processing function to handle data pipeline, model training, and model exchange. Since each FedEdge node provides some service in addition to consuming a service, EPR is designed to function in a dual role such as a server and client. In the client role, the FedEdge node can send registration requests, upload local models, and reply to the aggregator’s status query. On the other hand, it accepts requests from the aggregator to preprocess the dataset, launch training, and receive a global model in the server role. To adopt a synergy in FedEdge nodes, we proposed a communication protocol. The general idea of the FedEdge communication protocol (FedEdge COMM) is to define the messages that should be exchanged between the aggregator and worker nodes. Besides, the protocol also defines the status flags that should be set by each node based on the role type and its current status. The communication flow of our FedEdge COMM is shown in Figure \ref{fig:comapi}. Following the FedEdge COMM, the aggregator node first initializes the port for accepting connections from the worker. Besides, the connection state tracker is initialized to keep track of the nodes communicating with the aggregator. After this phase, worker nodes register their IP address and device ID to the worker registry within the connection state tracker module. Once the aggregator has received the training resources, it will initialize the global model and share it with all the workers within the worker registry. If the workers successfully receive the model, a notification will be sent to the aggregator. If the aggregator determines that it has the required number of workers for the training phase, the training request is dispatched to all workers with the batch size and number of rounds. On receiving training requests, the worker initiates a training round and updates the local model to the aggregator at the end of the training.

\subsubsection*{FL-Transport Layer}
FedEdge nodes on Wireless multi-hop networks are prone to experiencing dynamic network conditions that might cause unreliable transport layer operation. With modularity at the core of the FedEdge framework design, we let programmers freely define the underlying transport layer mechanism for the FedEdge COM API. While the EPR submodule exposes FedEdge Compute and dataset layers’ functions, these APIs are accessible based on the transport layer of FL (FL-Transport). With service-based architecture as the design principle behind FL-transport’s transport layer, we have adopted two communication mechanisms (1) HTTP- REST API and (2) GRPC.

\textit{HTTP-REST API:} The widely used service-based infrastructure protocol, HTTP- REST API operates on the principle of standard request/response format. Our HTTP- TCP REST API is constructed on top of the TCP protocol stack. Therefore, we extend REST APIs programmability to configure the number of TCP streams, session control, and keep alive. To simplify the payload transport, we utilized JSON message format to transport the FL model. In addition to the FL model, the user can easily extend the API to add additional attributes such as the timestamp of the model and the total time take to complete training by defining new key/value pair before JSON encoding. To simplify the implementation efforts, we adopted a range of frameworks such as FastAPI \cite{fastapi},  Asyncio \cite{asyncio} and HTTPx \cite{httpx } for building the communication protocol. FastAPI framework facilitates application developers to implement REST API for functions that can be invoked by remote nodes using HTTP as the transport protocol. Asyncio and HTTPx enable the developers to implement asynchronous HTTP clients so that the aggregator can communicate with all workers within the cluster concurrently. While server function within the aggregator and worker is built around FastAPI and HTTPx, client functions are built on top of AiOHTTP and Asyncio.

\textit{GRPC:} Google Remote Procedure Call \cite{grpc} is a high performance framework for calling remote methods by passing parameters alike local functions. The core idea of the GRPC framework is to define a service that will implement the interfaces which can be called by the remote entities to execute an operation. In FedEdge framework, we leveraged GRPC as the communication channel between the server and all workers. Besides, the dual message format is supported over the channel such using either JSON or Protocol buffers. While JSON encodes messages as texts, protocol buffers use a compiler to transcode structured data into serialized byte streams. In addition, there exists native support for asynchronous execution, HTTP2, and data compression, which significantly reduce the overall traffic volume in wireless multi-hop FL.

\subsection{Federated Networking}	
In the previous section, we detailed our design and implementation of federated computing. The key objective of our work is to improve the convergence time of federated learning systems by optimizing communication delay over multi-hop wireless networks. To tame the network latency  and to implement reinforcement learning routing module, first we need a platform that enables visibility of per-packet networking statistics (such as delay) for RL training. In addition, we need to realize distributed and programmable network control so that the MA-RL policies can be learned, deployed, and executed in a real-time fashion. To satisfy the above two requirements, we developed a federated networking subsystem by enhancing and customizing WiNOS, a distributed wireless network operating system proposed in our previous work \cite{infotelm}. The federated networking subsystem is composed of three layers (1) Dataplane, (2)  Network Core services, and (3) RL Applications. Compared with WiNOS, the new enhancements include the redesigned dataplane and upgraded core services to support multi-radio networking, customized in-band telemetry processing to support federated networking, and the new RL application with domain-specific action space refining. The details of some important components of federated networking system are introduced as below.

\subsubsection{\textbf{Telemetry-enabled Dataplane}} The crucial function of dataplane is to forward packets in-line with the native Linux wireless MAC80211 network stack and to provide programming primitives to control packet forwarding. To enable programmable packet forwarding on wireless multi-hop networks, we leveraged Openflow-based Datapath to send and receive data packets. Our datapath is developed using Openflow Software switch, namely Ofsoftswitch13 \cite{userswitch}. Dataplane functionality is pivotal for realizing AI-enabled forwarding schemes. Nowadays, dataplanes are not only designed to handle packet forwarding, but they also gather vast amount of data for network monitoring using SNMP, sFLOW, Collectd, and many more. However, existing solutions do not support actionable data sampling or measurement schemes that can be rapidly exploited for routing schemes with delay minimization as the objective. In addition, they require additional channel resources and fail to capture real-time delay of packets. 

With an AI-enabled platform as the core of our system design, we proposed and developed a real-time in-band network telemetry module. The primary objective of our telemetry solution is to collect real-time experience of packets, such as per-hop delay over each traversing link or end-to-end link, in a cost-efficient manner. Towards this end, we have developed and implemented  a  distributed in-band  telemetry  system \cite{infotelm}, where  each  router  runs  its  own  telemetry  module built  on  the  top  of  OpenFlow  processing  pipeline. Our in-band  telemetry  system  consists of a new telemetry packet header, two new packet matching actions (i.e., PUSHINTL  and POPINTL) and the telemetry processor. To assist AI-enabled federated networking, we reconfigure the telemetry processor so that whenever a FL packet passes through the router, the router will recognize such packet and insert a timestamp (i.e., the time instance when FL packet arrives at the router) into the telemetry packet header. Then, by using the PUSHINTL action, the router performs packet encapsulation by adding the telemetry packet header into the FL packet. After receiving such FL packet, the next-hop router decapsulates the FL packet via POPINTL action and retrieves the timestamp. The difference between the timestamps of the sending router and receiving router is the one-hop packet delivery delay. The key advantage of our in-band telemetry is that the routers are able to use data packets to carry measurement data with minimum cost, where the measurement data or experience are the keys for training RL agents. In this work, the in-band  telemetry system is tested and optimized so that the extra delay induced by the telemetry processing operations is negligible.



Last, wireless networks have another dimension of control on PHY, such as the channel and power, that may significantly affect its overall performance. We have extended programmability to control PHY using NetLink interface from the controller. Each RF hardware on the router is attached as a virtual port on the datapath and link layer discovery such as neighbor peering is handled by MAC80211 stack.

 \begin{figure}[h!]
	\centering
	\includegraphics[width=0.9\linewidth]{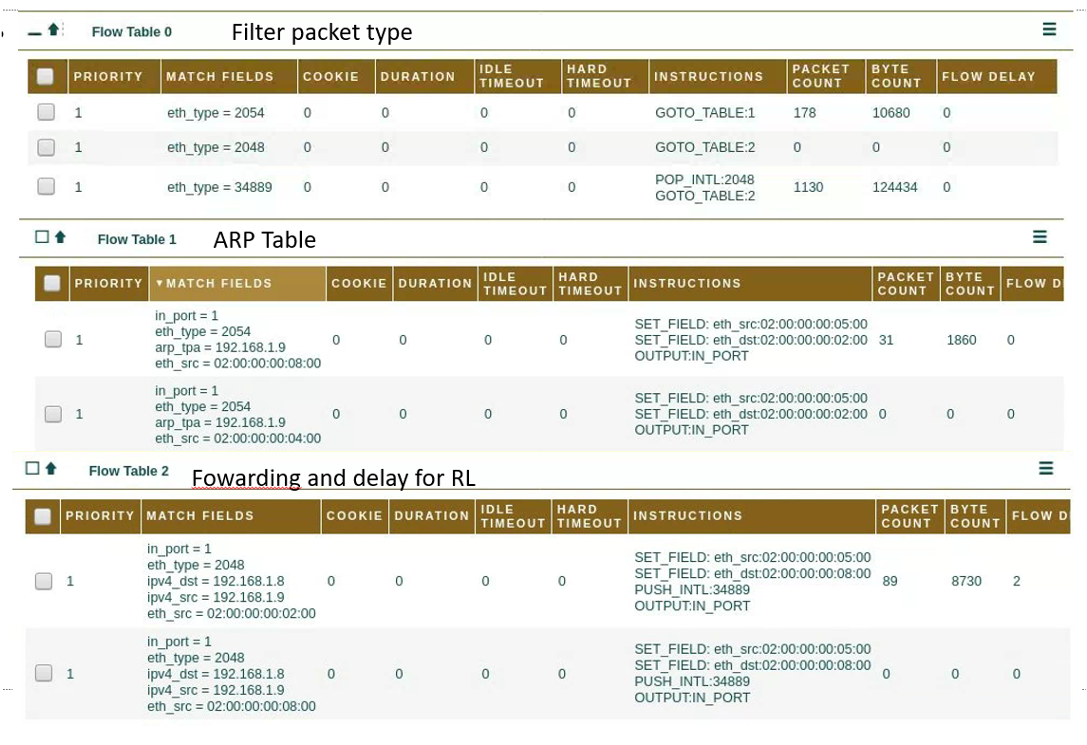}
	\caption{MultiFlow Table}
	\label{fig:multiflow}
\end{figure}
\subsubsection{\textbf{Network Core Services}} OpenFlow-enabled controller provides core services for interacting with the datapath and to develop network applications for orchestrating the packet handling behavior. Our core services include  OpenFlow Manager based on RYU controller, telemetry manager, network state, and telemetry database based on MangoDB \cite{mangodb}, and radio interface manager based on NetLink library. Openflow manager is responsible for providing the required APIs or handlers to monitor and control the datapath in realtime by adding/removing entries into OpenFlow table

Standard packet forwarding behavior such as receiving and forwarding packets over ports can be realized using a single flow table. However, the complexity of the flow table increases exponentially when handling telemetry packets due to the nature of sequential processing of flow table instructions. Hence, we leveraged multi-flow table based flow instructions for handling packet forwarding as shown in Figure \ref{fig:multiflow}. First, table-0 instructions will identify the presence of telemetry by matching the ether\_type and then relevant action is performed. Second, table-1 will handles ARP requests/reply and finally table-2 flow instructions perform the actions for forwarding the packets to the output port. Telemetry manager instructs and gathers the packet flow monitoring metrics from the telemetry enabled datapath. Our network state database provides RPC based interfaces for data access within the kernel layer and also provides access interfaces via Northbound API’s for network applications, such as reinforcement learning based routing algorithms. 


\textit{Line-speed Action-state Value Estimation:} The key challenge to implement reinforcement routing algorithms is how to estimate the action-state values (i.e., Q values) without inducing so much control overhead. In particular, estimating Q values relies on the measurement of per-hop per-packet delay as shown in eq. \eqref{EWA}. Directly requesting the delay information from the neighboring router could introduce significant overhead to the bandwidth-limited wireless channel. Therefore, it is necessary to redesign the way of exchanging information among neighbors. We design the line-speed Q value estimation for each router $i$, which aims to realize Q estimation at the line speed, i.e., the speed at which packets come in the router. The local $Q_i$ estimation of router $i$ is directly coming from its next-hop neighbor.  The motivation of such design is based on the fact that the action-state value $Q_i$ of router $i$ is estimated based on the per-packet reward (per-packet delay) $r_i$ and the action-state value $Q_{i + 1}$ of the next-hop router $i+1$. Both $Q_{i+1}$ and per-packet reward $r_i$ is immediately available at next-hop router $i+1$ where $r_i$ is obtained via the in-band telemetry introduced above. Therefore, it is more cost-effective to let next-hop router $i+1$ estimate the action-state value $Q_i$ of the current router $i$ via exponential moving average. The estimated action-state value $E(r_i + Q_{i + 1})$ is sent back by the next-hop router $i+1$ to current router $i$ periodically (e.g., every five seconds).  Such a scheme allows the action-state value to be updated at the line speed, while orderly reducing the control overhead.
 
 \begin{figure}[h!]
	\centering
	\includegraphics[width=1\linewidth]{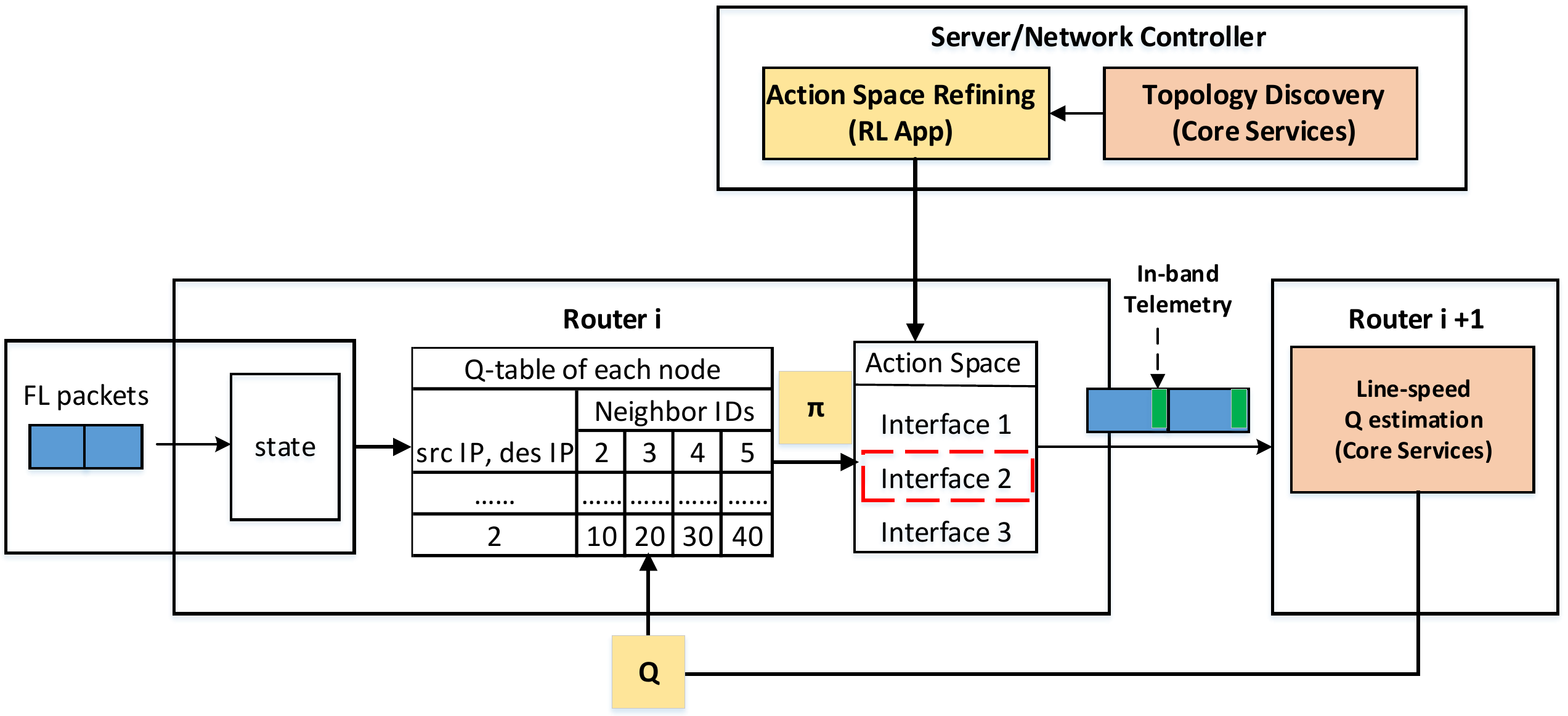}
	\caption{RL Application with tabular Q estimation}
	\label{fig:router1}
\end{figure}

\subsubsection{\textbf{RL Application}} RL routing solution is developed as a network application that can access and pass messages using core services REST API. For example, to implement an Actor/Critic RL algorithm, the first step is to build the Q-table based on the local topology information retrieved from the network state database as shown in Figure \ref{fig:router1}. Following this, we can initiate the Q-table using the estimated Q-values stored in the database. Besides the Q table, we can also adopt neural networks for non-linear Q value approximations especially when more network states are utilized, such as traffic matrix and queue lengths. We will incorporate Q approximation neural networks function in the future work. Actor, on the other hand, disseminates the routing control decisions following the critics and policy $\pi$ to the OpenFlow manager.  OpenFlow manager translates the RL decisions into actionable OpenFlow datapath instructions for orchestrating the packet forwarding behavior. To accelerate the RL policy convergence, the loop-free action space is calculated by the action space refining application (introduced in section III.C). This application is running on the network controller, which has the global network topology readily available via the topology discovery module. Our topology discovery module can operate in either a centralized or distributed manner. The centralized approach directly uses the link layer discovery protocol (LLDP) that is initialized and coordinated by the network controller. For the distributed approach, each router discovers its one-hop neighbors via IEEE 802.11 local topology discovery scheme and the local typologies from all the routers are then aggregated by the network controller to form the global topology. It is worthy to note that both the network controller and wireless routers employ the same federated network subsystem shown in Figure~\ref{fig:FedEdge}, except that the network controller runs an additional action space refining application.

 \begin{figure}[h!]
	\centering
	\includegraphics[width=1\linewidth]{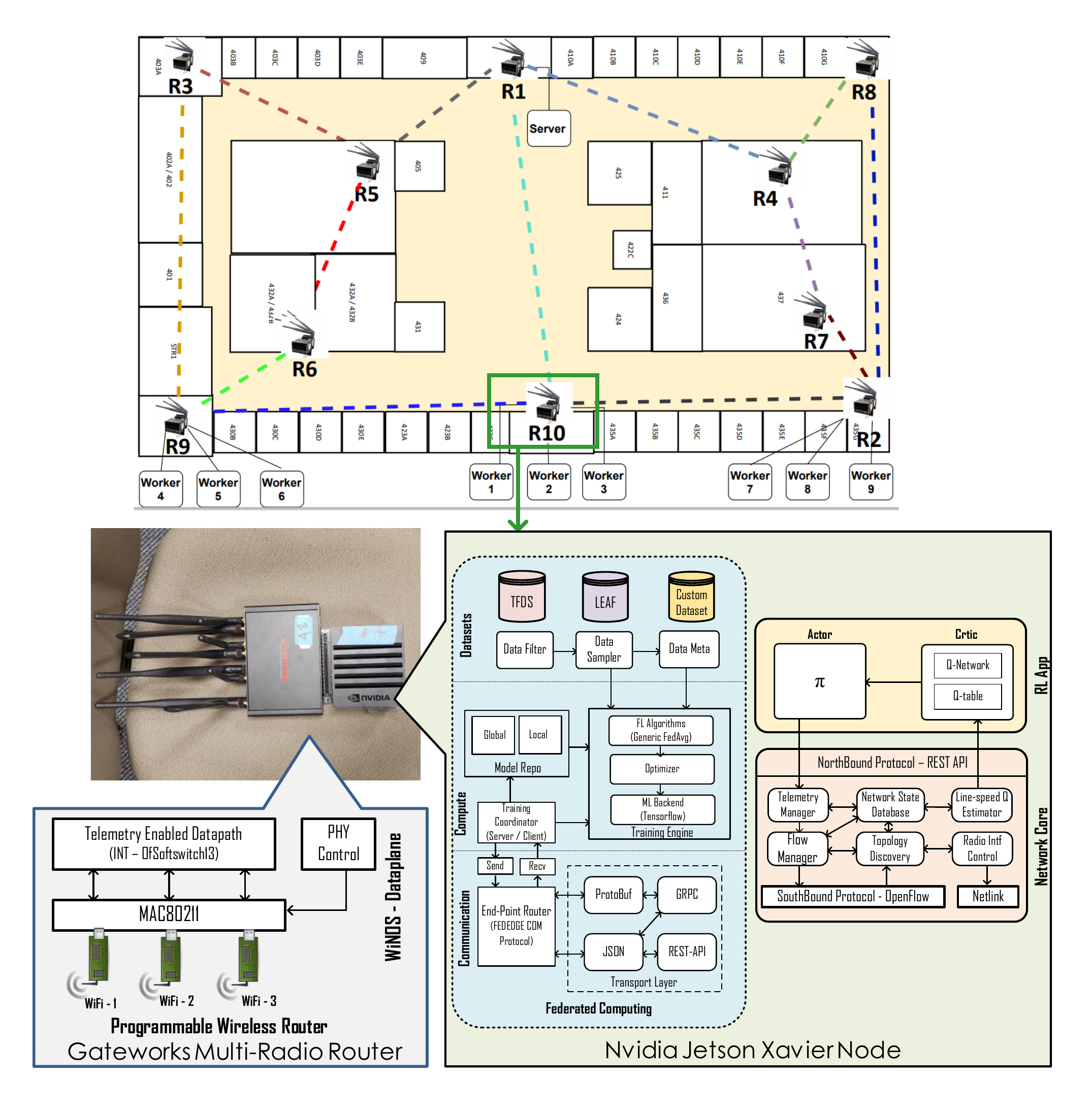}
	\caption{FedEdge - Testbed Topology and Node View}
	\label{fig:testbed_topology}
\end{figure}

\section{System Implementation}
\subsection{Overall Implementation}
We prototype our FedEdge system with a mesh topology as shown in Figure~\ref{fig:testbed_topology}. This testbed consists of 10 Nvidia Jetson Xavier nodes, each of which is connected to one Gateway 5400 multi-radio wireless router. Each Nvidia Jetson node serves as federated computing node which handles the federated learning training. In addition, the network core and RL-app of federated networking subsystem are also hosted on Nvidia Jetson node. On the other hand, Gateworks routers serve as federated networking node that only hosts a dataplane submodule. The dataplane is orchestrated using OpenFlow protocol by the network core services hosted on Nvidia jetson node. Such decoupled system design allows resource-rich Nvidia nodes to handle the computation-intensive FL operations and RL-based networking intelligence while keeping the operations of resource-limited routers simple and fast. 

\subsection{Federated Networking Subsystem Implementation}
Our multi-radio wireless federating networking nodes (Gateway routers) are off-the-shelf small-factor single-board computers which support Linux operating systems with multiple PCIe slots for adding wireless radio cards. In our testbed, we deployed Ubuntu 20.04 as the operating system and 3 x Compex WLE900VX-I wireless cards to enable multi-radio wireless nodes. Each wireless radio is set to operate on 5 Ghz channels and 20 Mhz channel width in 802.11ac operating mode with 15 dBm transmission power. As a result, each wireless router in our testbed can reach roughly 40 Mbps aggregated data rate from three radio cards. On top of the node operating system, we deploy a dataplane submodule that facilitates a software bridge with a programmable packet handling routine (i.e OpenFlow Flowtable). All three wireless cards were configured to operate on disjoint channels, and then they were added to OpenFlow bridge as OpenFlow ports. Since our router is embedded hardware with limited computation power and cpu cores, the timley availability of CPU processing cycle is essential for seamless performance. Hence, in our testbed, we explicitly define CPU cores for the OpenFlow process by using Linux \textit{Taskset} functionality.

 \begin{figure}[h!]
	\centering
	\includegraphics[width=1.0\linewidth]{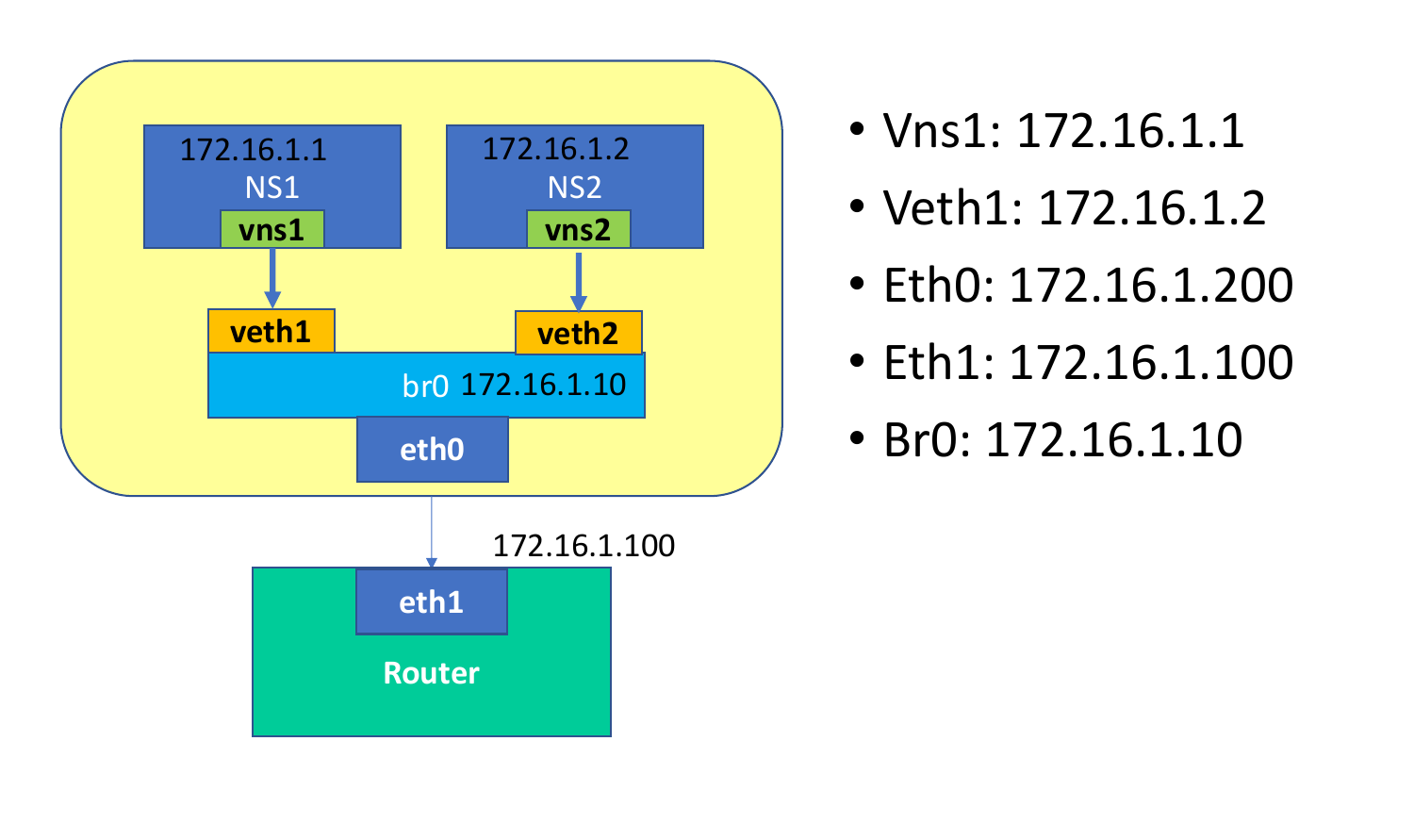}
	\caption{Federated Computation - Namespace Isolation}
	\label{fig:ns}
\end{figure}

\subsection{Federated Computing Subsystem Implementation}
A federated computing system is deployed in Nvidia Jetson Xavier node with has 16GB combined RAM for GPU and CPU. Jetson nodes come with Ubuntu 20.04 operating systems and TensorFlow packages as part of the hardware. In our testbed, each jetson node can host different number of FL worker nodes. Since the primary goal of our experiment is to study the impact of wireless networking on FL, we enabled isolation only at the network level using network namespace. The main advantage of such approach is that, within the single hardware we can deploy multiple workers with isolated TCP/IP layer. Figure~\ref{fig:ns} shows the schematic of the namespace virtualizated network for worker on each jetson node. On each  jetson node, we create a virtual bridge using Linux brigde-utils and then we create namespaces for each worker.  Following that, the interfaces from the namespace is added to the newly created bridge using virtual ethernet pairing. At this point, all workers within each jetson node can communicate independently. To facilitate external connectivity, nodes master ethernet interface is also added to the same bridge. Finally, each worker can be launched from their respective namespace by executing the API scripts.

\section{Experimental Evaluation}\label{sec:eval}
We conduct extensive experiments to evaluate the effectiveness of our proposed networked-accelerated FL system on our physical testbed. We will compare the performance of our proposed RL-based federated networking with widely-adopted production-grade wireless networking protocol BATMAN-ADV \cite{batman} under a variety of settings by varying the number of workers, the percentage of stragglers, and  worker location distributions.


\subsection{Experiment Setup}
\textbf{Model and Dataset:} 
Our experiments consider image classification tasks on FEMINIST and CIFAR-10 datasets. Two different models are used in the training experiments—a shallow two layers of CNN model and MobileNet, whose weights are updated using federated learning.
\begin{itemize}

\item FEMNIST CNN:
We first use FEMNIST, the federated version of MNIST \cite{mnist} on the LEAF\cite{leaf} character recognition task, where LEAF is a benchmarking framework for federated learning. FEMNIST consists of handwritten digits (10), uppercase (26), and lowercase (26) letters leading to a total of 62 classes with each image having 28$\times$28 pixels. The whole dataset is partitioned into 3550 data portions/users with Non-IID data distribution. In our experiments, we sub-sampled the dataset by 0.02\%, yielding 71 users.

Initially, we evenly distribute these users among three edge routers R9, R10, and R2 as shown in Figure~\ref{fig:testbed_topology}. For each router, we assign three active workers with each active worker consisting of 7--8  users, and one of the 7--8 users will become the active worker at each global round/epoch of federated training. We employ a convolutional neural Network (CNN) model during testing. The CNN model has two convolution layers, with 32 and 64 filters respectively. Each convolutional layer was followed by a 2x2 max pooling layer. The convolutions were followed by a fully connected layer with 128 units with ReLU activation. A final fully connected layer with softmax activation was used as the final output layer. The model has a size of 5.8 Mbytes. 

\item CIFAR-10 MobileNet:
To demonstrate the feasibility of a real-world scenario, we introduce the CIFAR-10 dataset and the MobileNet model. CIFAR-10 has 10 classes, 50,000 training samples, and 10,000 testing samples. We used the Dirichlet distribution Dir($\beta$) to build Non-IID heterogeneous partitions for all workers. The value of beta is 0.5, determines the degree of heterogeneity. To train the CIFAR-10 dataset, we use a MobileNet model, a class of efficient network architectures for low power computing devices such as the Nvidia Jetson Xavier platform. To deploy multiple models in resource-constrained hardware, we reduced the width size of the model to be thinner with a width multiplier ($\alpha$) of 0.5, and set the input resolution of the network to 224. Our model has a size of 7 Mbytes.

\end{itemize}

\textbf{Baseline Federated Networking Protocol:} To compare the performance of our proposed RL based routing for federated learning, we chose the state of the art mesh routing protocol, BATMAN-Adv \cite{batman} as the baseline. BATMAN-adv is implemented as a layer 2 proactive routing protocol based on distance vector and radio link based reliability as the routing metric. In addition, each node only maintains route information to the next node by which the final destination can be reached. Since each node only requires next hop information towards the destination node, global exchange of routing information is not necessary. From the aforementioned operation of BATMAN-adv, we identified it as the best candidate for comparison as the operation of our RL routing scheme also utilizes only next hop nodes information. Moreover, to the best our knowledge, BATMAN-Adv is the only multi-radio mesh routing protocol that works out of the box on Linux systems as it is embedded within the Linux kernel for optimized operation. During the experiments, we directly use the production-grade BATMAN-Adv protocol provided by Linux system.

\textbf{RL-based Federated Networking Protocols:} We study two RL-based networking algorithms including on-policy greedy algorithm and on-policy softmax algorithm. As introduced in Section III, on-policy greedy algorithm uses greedy policy for both target policy and behavior policy. For on-policy softmax, both target policy and behavior policy use  softmax-greedy policy defined in eq. \eqref{softmax}.

\textbf{Hyperparameters:} For FL, we use the batch size of 100 and learning rate of 0.1. For MA-RL, we use the learning rate of 0.7 for both RL approaches and temperature $\tau$ is set to be 2 for on-policy softmax. We did hyperparameter search for $\tau$ and it shows that different values of $\tau$ do not lead to significantly different performance.

\subsection{Main Results}
\begin{figure}[]
\captionsetup{singlelinecheck = false, justification=justified}

	\subfigure[Iteration loss convergence]{
    \includegraphics[width=0.43\linewidth]{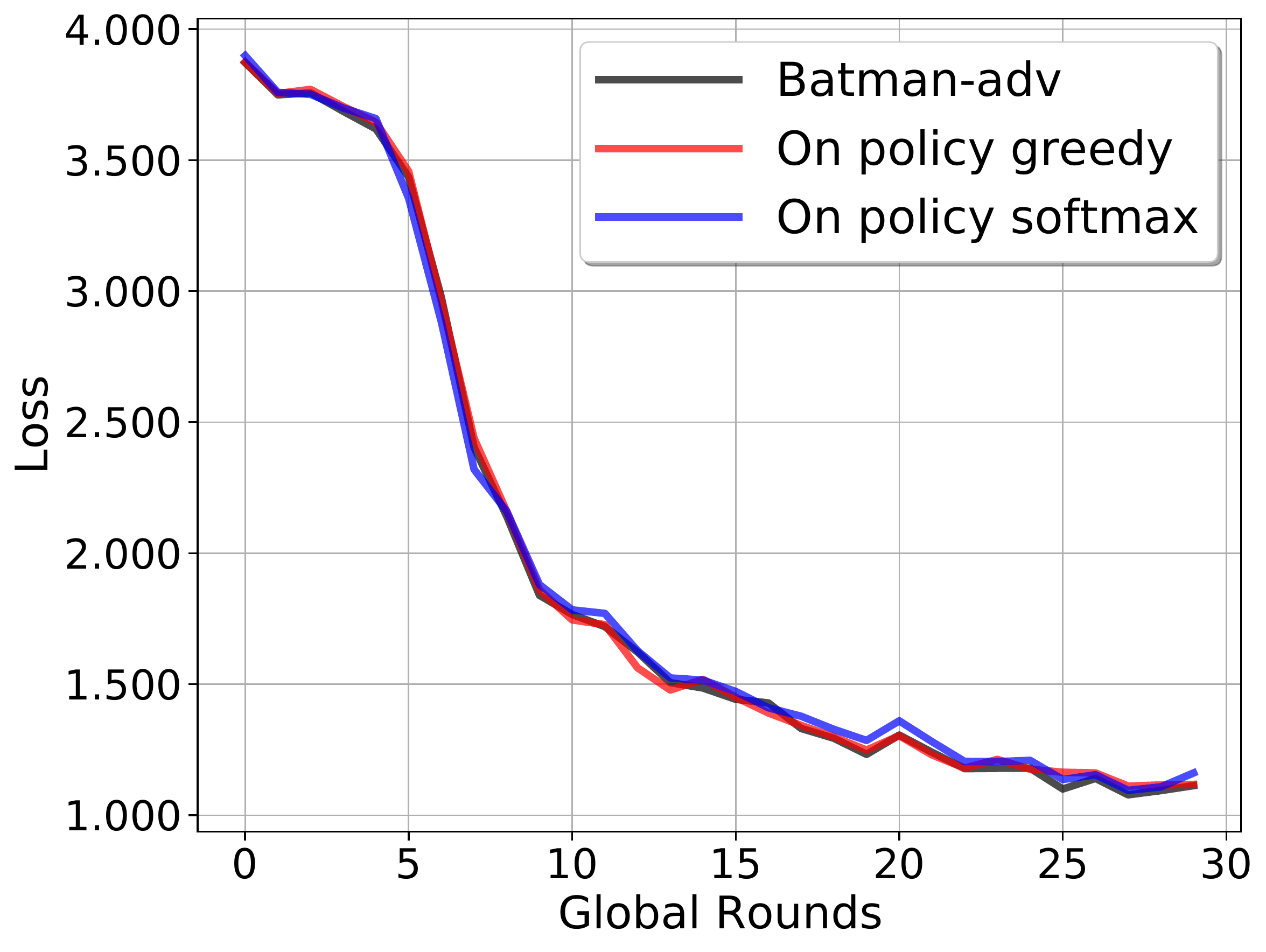} 	}
	\subfigure[Wall-clock loss convergence]{
    \includegraphics[width=0.43\linewidth]{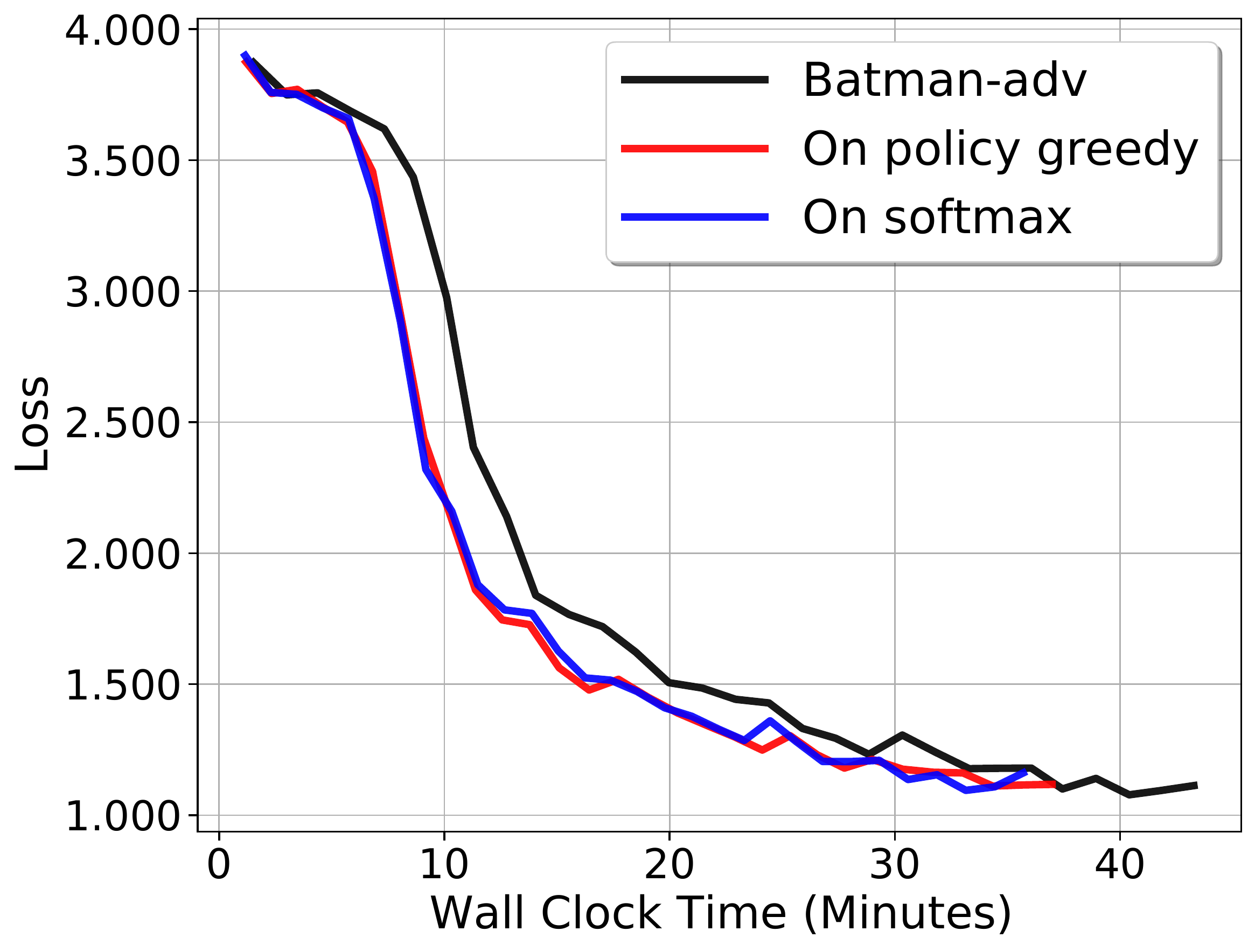} 	}
	\vspace{-0.3cm}
	\caption{Loss convergence comparison of BATMAN-Adv, on-policy greedy, and on-policy softmax with 9 workers} 
 	\label{fig:acc}
  	\vspace{-0.2cm}
	\subfigure[Iteration accuracy convergence]{
 	\includegraphics[width=0.43\linewidth]{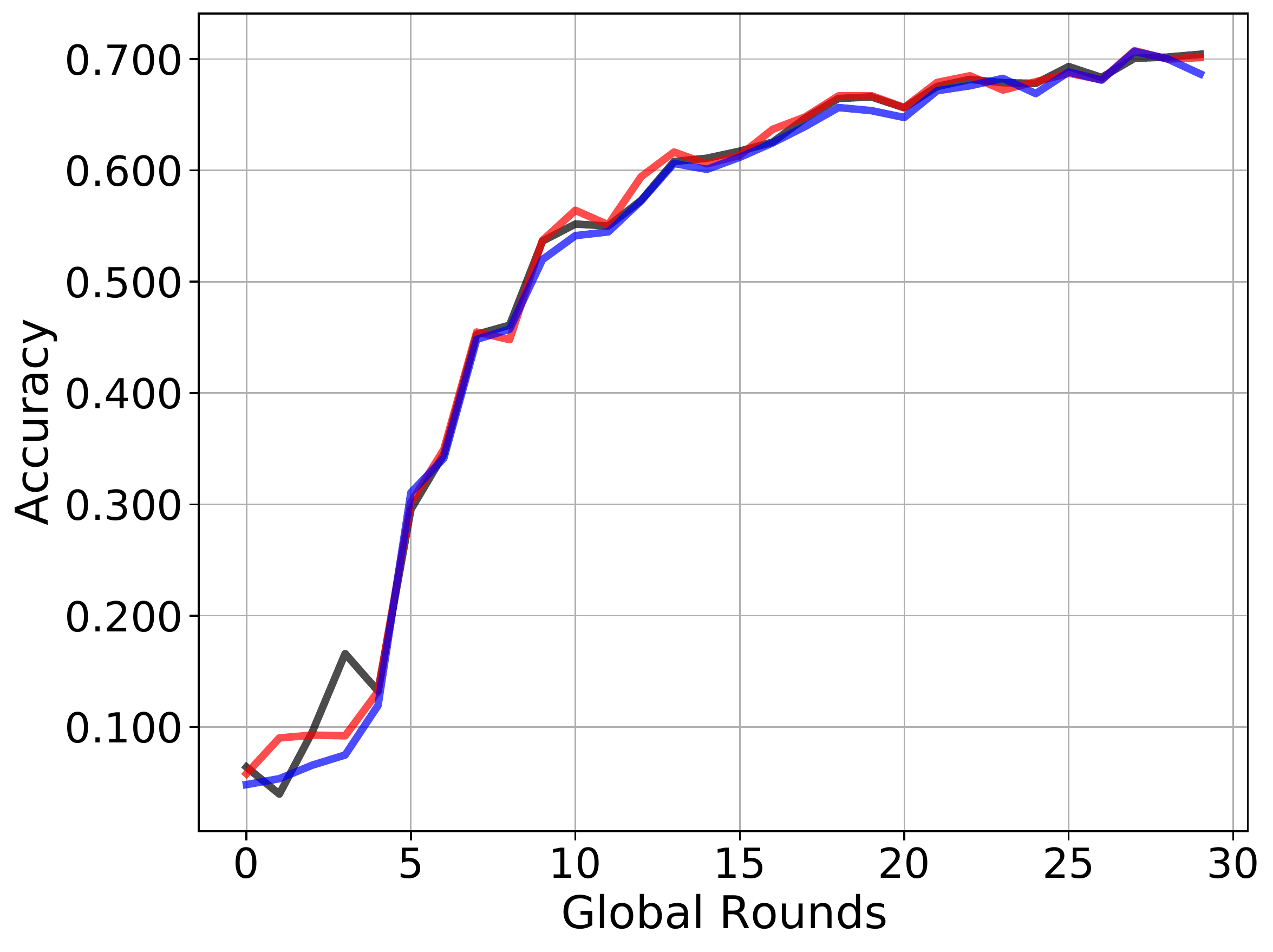} 
 	}
	\subfigure[Wall-clock accuracy convergence]{
 	\includegraphics[width=0.43\linewidth]{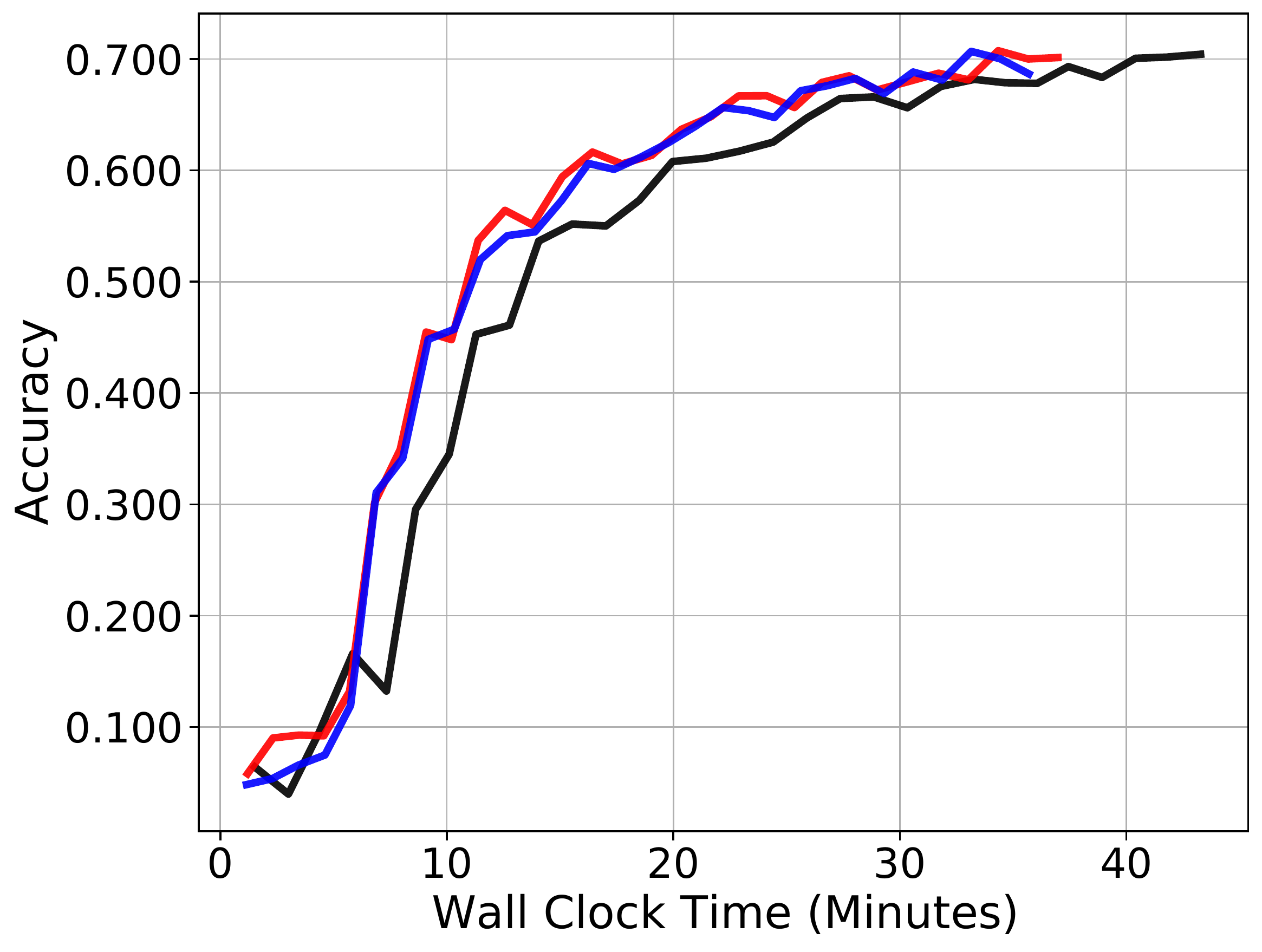} 
	}
	\caption{LEAF and 2-CNN: Validation Accuracy convergence comparison of BATMAN-Adv, On-policy greedy, and On-policy softmax with 9 workers}
 	\label{fig:total_time}
 	\vspace{-0.5cm}
\end{figure}
\subsubsection*{\textbf{FL iteration and wall-clock convergence}} As shown in Figure~\ref{fig:acc} and \ref{fig:total_time}, all three federated networking protocols lead to the same iteration convergence performance in the sense that they achieve the same loss or validation accuracy after running the same number of epochs. This is as expected because they use the same underlying federated training algorithm. However, RL-based federated networking protocols can achieve much better wall-clock convergence performance, compared with the baseline protocol. This is because RL algorithms can minimize the per-epoch duration by learning the delay-minimum forwarding paths for model exchange between the server and workers.

 \begin{figure}[]
 \captionsetup{singlelinecheck = false, justification=justified}

	\centering
	\includegraphics[width=0.8\linewidth]{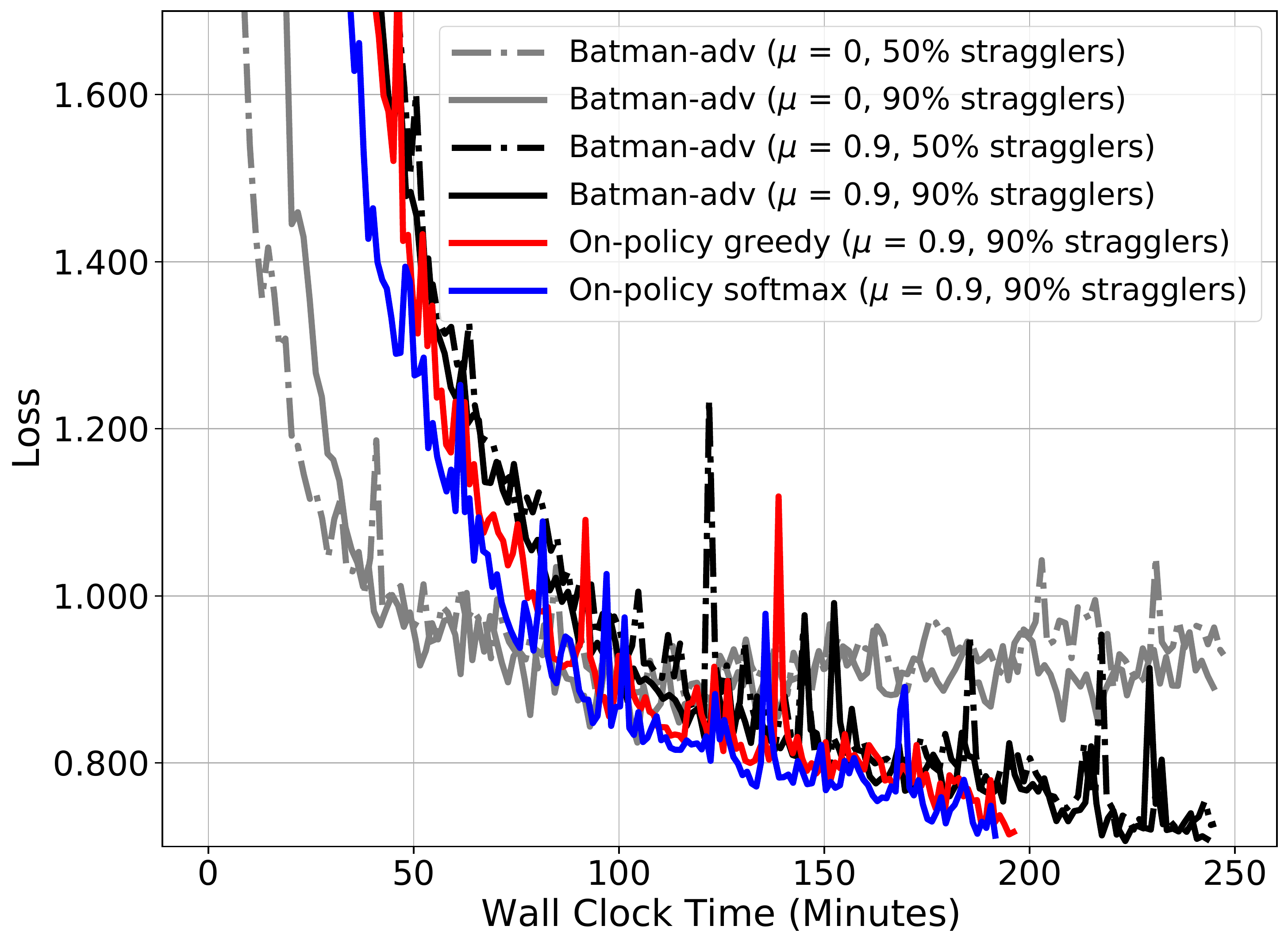}
	\caption{LEAF and 2-CNN: Loss Convergence Time after 170 global rounds under different routing protocols } 
	\label{fig:fedprox}
\end{figure}

 \begin{figure}[h!]
 \captionsetup{singlelinecheck = false, justification=justified}

	\centering
	\includegraphics[width=0.8\linewidth]{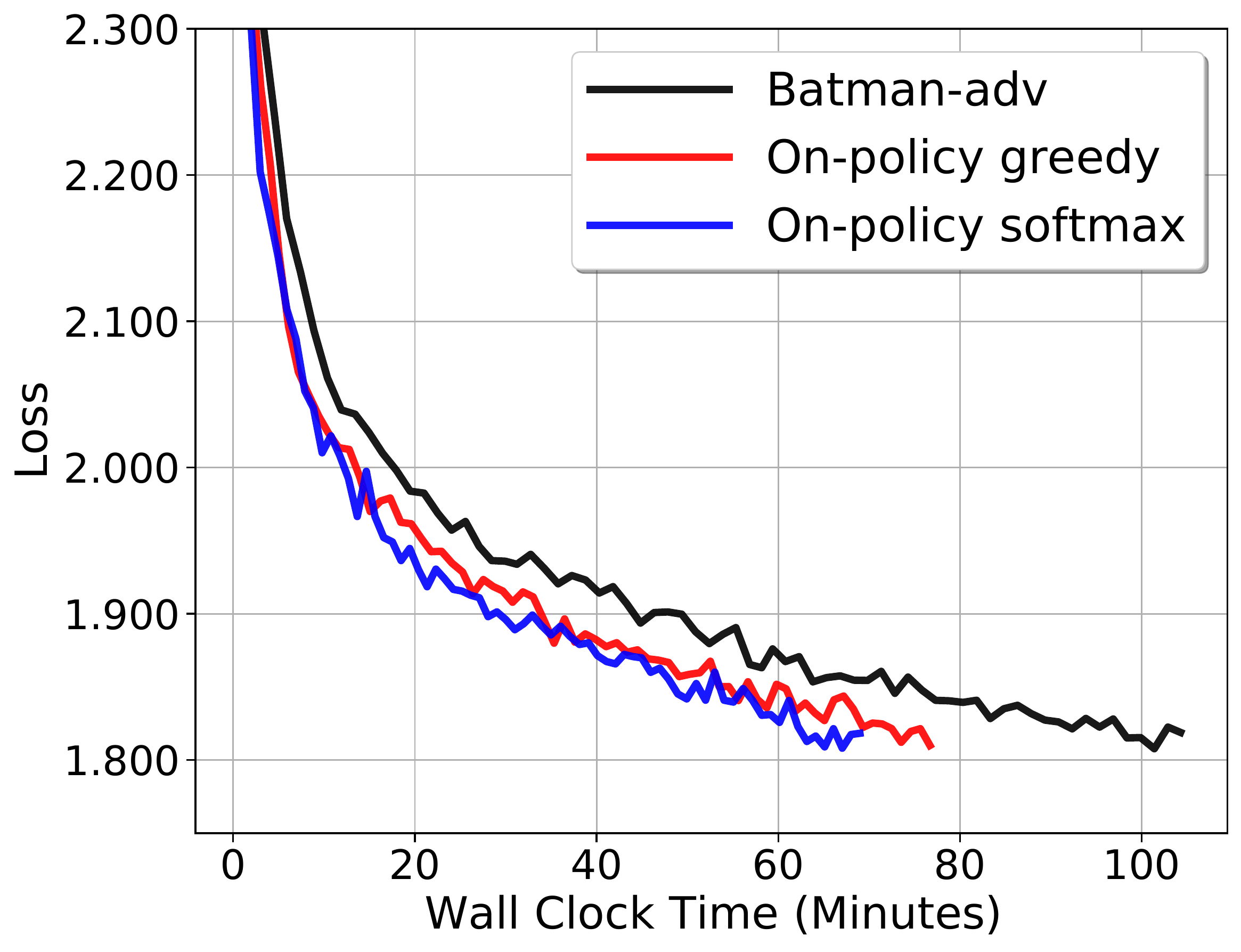}
	\caption{CIFAR-10 and MobileNet: Loss Convergence Time after 70 global rounds under different routing protocols } 
	\label{fig:mobilenet_loss}
\end{figure}

\subsubsection*{\textbf{Effect of stragglers on FL convergence}} In this evaluation, we show that by performing regularized model updates, 
FL convergence performance is improved in presence of heterogeneous workers under both RL-based federated networking and baseline approach. We study the effect of stragglers (or computationally-slow workers) that use heterogeneous or different numbers of local epochs, which is smaller than the non-stragglers that use the maximum number of local epochs. Such heterogeneous setup will lead to model divergence because of varying number of local updates by workers. Then, the straggler effect is evident from the noisy updates when the model is unregularized (i.e., $\mu = 0$). When the local model updates are regularized (i.e.,$\mu > 0$), the convergence is less noisy and prevents the model divergence. We trained the FL model under different straggler percentages including 50\% and 90\%, It is shown in Figure~\ref{fig:fedprox} that less percentage of stragglers leads to less noisy local model updates and thus leads to faster convergence especially at initial training epochs. In addition, wall clock time is significantly reduced by 50 minutes when RL algorithms is used for routing even in presence of stragglers. By applying regularized SGD, better convergence performance can be observed when the number of training epochs increases, even though at the initial training stage, regularized SGD converges slower than the classic SGD.  

\subsubsection*{\textbf{Results of Loss Convergence on CIFAR-10 and MobileNet}}

Figure~\ref{fig:mobilenet_loss} presents the performance comparison of CIFAR-10 and MobileNet with different routing algorithms in terms of loss convergence and wall clock convergence time. The results become more promising with a larger model size, which lead to higher FL traffic in the network.
Both RL routing reach the same loss convergence by around 70 and 79 minutes, respectively, approximately 35 minutes faster than the BATMAN-Adv baseline routing, which takes almost 110 minutes to achieve the same loss convergence.

 \begin{figure}[h!]
 \captionsetup{singlelinecheck = false, justification=justified}

	\centering
	\includegraphics[width=0.8\linewidth]{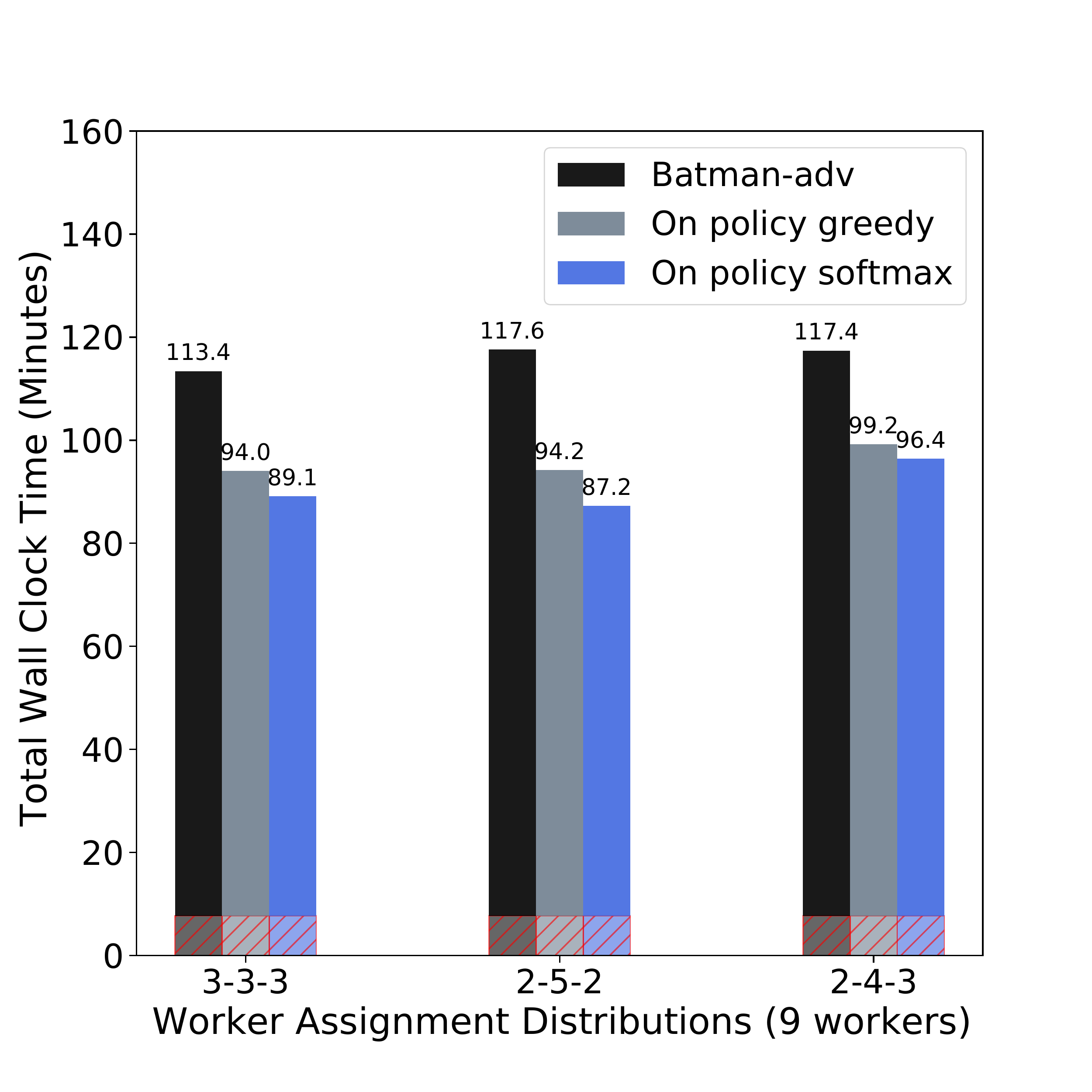}
	\caption{LEAF and 2-CNN: Total convergence time comparison of Batman-adv routing (black), On-policy greedy (grey), On-policy softmax (light blue), and Computation time (red hatched) under different worker location distributions after 80 global rounds. } 
	\label{fig:workerdis}

\end{figure}

\subsubsection*{\textbf{Impact of worker location distribution}} In Fig \ref{fig:workerdis}, we investigate how worker location distribution affects the FL convergence performance. We study the total FL convergence time, FL computing time, and FL networking time by varying the number of workers that are connected to the three edge routers (R9, R10, R2). We study three node distributions (3-3-3, 2-5-2, 2-4-3) with a total number of 9 workers.  It is evident that the RL-based federated networking can consistently outperform the baseline networking protocol under different node distributions and achieve up to 25\% convergence speedup, compared with the baseline. Moreover, when the network becomes congested, RL-based federated networking protocols lead to a higher performance gain because they can learn to maximize the network resource utilization to better distribute FL flows among all available forwarding paths. This advantage is shown under 2-5-2 worker distribution, where the router R10 needs to serve 5 workers, which induces higher FL traffic volume and a higher level of network congestion around router R10. In this case, on-policy softmax policy leads to the 25\% speedup, which is the maximum one among the three node distributions. Regarding RL-based approaches, on-policy softmax outperforms on-policy greedy for all three cases. This is due to the fact that on-policy softmax can proportionally distribute the traffic flows among the available forwarding paths according to the E2E delay of each path. Such approach could be more effective to distribute the traffic loads. In addition, it is observed that the majority of total run time came from communication time while the computation time (around 8 minutes) only contributes to a small portion of the total training time. Therefore, optimizing the federated networking performance is very beneficial to accelerate FL convergence in multi-hop edge computing networks.

 \begin{figure}[h!]
 \captionsetup{singlelinecheck = false, justification=justified}

	\centering
	\includegraphics[width=0.8\linewidth]{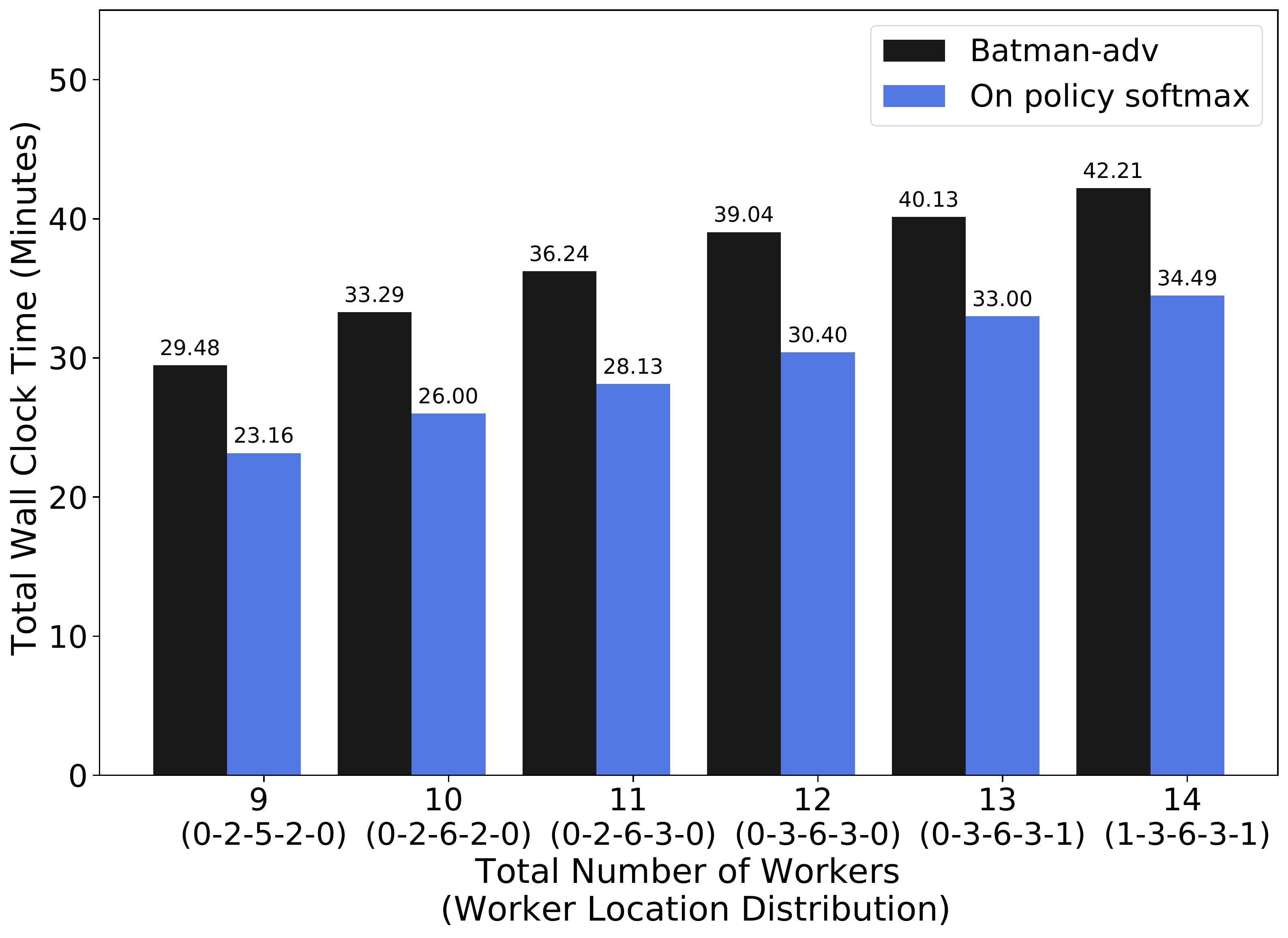}
	\caption{LEAF and 2-CNN: Total convergence time comparison of Batman-adv routing (black), On-policy softmax (light blue), by varying total number of workers and location after 20 global rounds. } 
	\label{fig:scale_cnn}
\end{figure}

 \begin{figure}[h!]
 \captionsetup{singlelinecheck = false, justification=justified}

	\centering
	\includegraphics[width=0.8\linewidth]{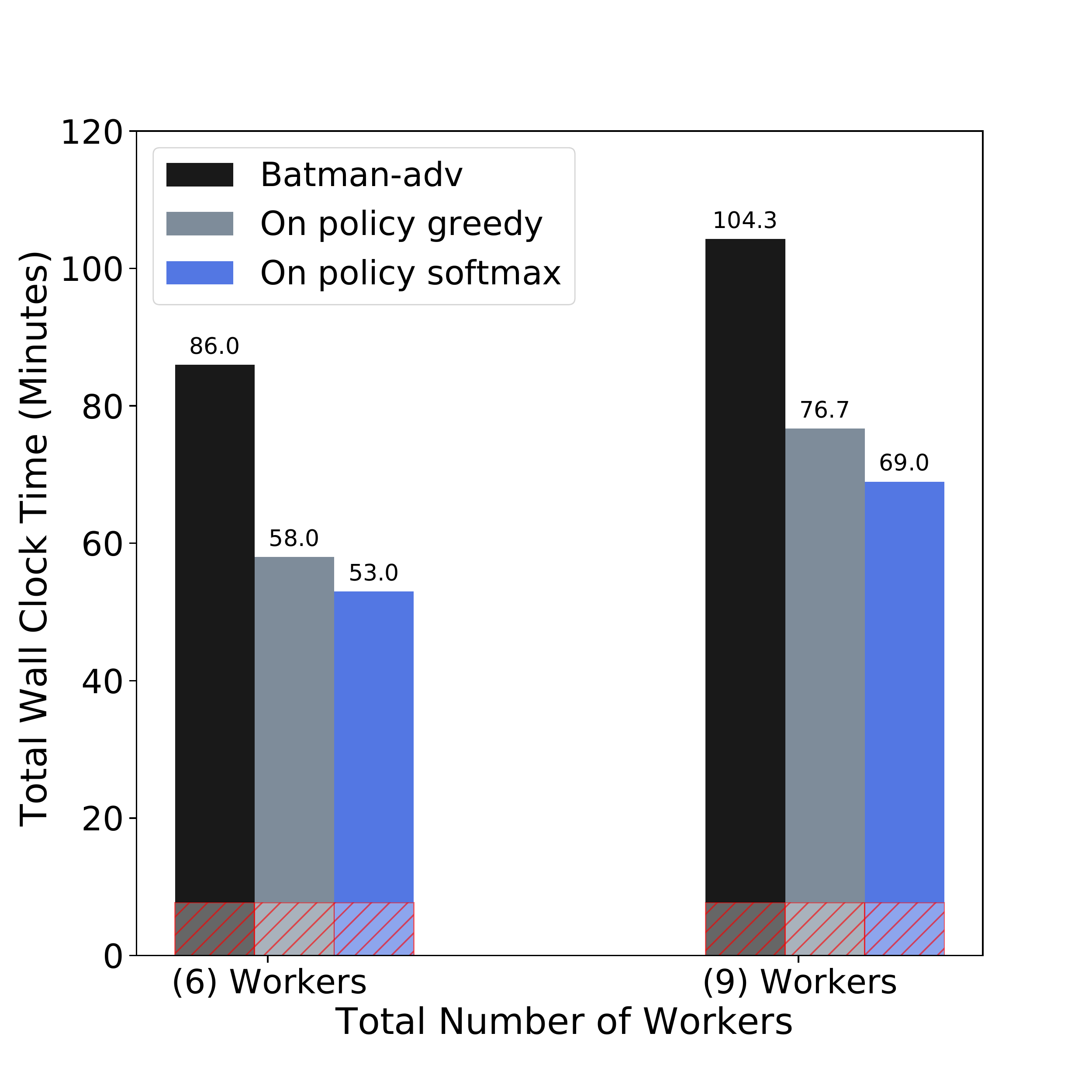}
	\caption{CIFAR-10 and MobileNet: Total convergence time comparison of Batman-adv routing (black), On-policy softmax (light blue), by varying total number of workers after 70 global rounds. } 
	\label{fig:mobilenet_time}
\end{figure}

\subsubsection*{\textbf{Scalability Analysis}}
In Fig. \ref{fig:scale_cnn}, we evaluate the FL convergence time by varying the number of workers attached to five edge routers (R9, R10, R2, R3, R8) with a total number of workers (9 , 10, 11, 12, 13, 14). As the number of workers increases, the convergence time of both RL-based approach and baseline protocol increases. This is because as more workers participate, the total FL traffic volume injected into the network increases and gradually approach the maximum network capacity. This leads to prolonged E2E delay and model training time. However, as shown in Fig. \ref{fig:scale_cnn},  RL-based approach keeps outperforming the baseline scheme consistently and reduce the total time by 23\%. That is, it is able to learn the delay-minimum forwarding paths even if the network becomes congested.

We only experimented with 6 and 9 workers for CIFAR10 and MobileNet because of resource-constrained hardware, as the Nividia Jetson node can only support up to 3 workers per device. As shown in Fig. \ref{fig:mobilenet_time}, we observed the similar trend as the number of increases, the convergence time of all routing algorithms also increase. However, the RL-based method continues to outpace the baseline scheme. Consistently, resulting in a 30\% reduction in overall time and achieved the same level of loss and a final accuracy of  68\%.

\section{Conclusion}\label{sec:conclusion}
This paper proposes network-accelerated FL over wireless edge by optimizing the multi-hop federated networking performance. We first formulate the FL convergence optimization problem as a  Markov decision process (MDP).  To solve such MDP,  we propose the multi-agent reinforcement learning (MA-RL) algorithm along with loop-free action space refining schemes so that  the delay-minimum forwarding paths are learned to minimize the model exchange latency between edge workers and the aggregator. To  fast prototype, deploy, and evaluate our proposed FL solutions, we develop FedEdge, which is the first experimental framework in the literature for FL over multi-hop wireless edge computing networks. Moreover, we deploy and implement a physical experimental testbed on the top of the widely adopted Linux wireless routers and ML computing nodes. Such testbed can provide valuable insights into the practical performance of FL in the field. Finally, our experimentation results show that our RL-empowered network-accelerated FL system can significantly improve FL convergence speed, compared to the FL systems enabled by the production-grade commercially-available wireless networking protocol, BATMAN-Adv.

\bibliographystyle{IEEEtran}

\bibliography{bib/bib-lee-marl,bib/FL_pu,bib/bib-Chen,bib/UW_Pu,bib/SAS_Pu,IEEEabrv,bib/bib-prabhu}

\end{document}